\documentclass[reqno,12pt]{amsart}

\usepackage{enumerate}
\usepackage[margin=1in]{geometry}
\usepackage{enumitem}
\usepackage{ifpdf}
\usepackage{amsmath}
\usepackage{amsfonts}
\usepackage{amssymb}
\usepackage[foot]{amsaddr}
\usepackage{amsthm}
\usepackage{mathrsfs}
\usepackage{amsrefs}
\usepackage{cite}
\usepackage[titletoc,title]{appendix}
\usepackage[dotinlabels]{titletoc}
\usepackage[all,cmtip]{xy}
\usepackage{nicefrac}
\usepackage{float}
\usepackage{verbatim}
\usepackage[multiple]{footmisc}
\usepackage{mathdots}
\usepackage{dcolumn}
\usepackage{tabu}
\usepackage[hang,small,bf]{caption}    
\usepackage[overload]{textcase} 
\usepackage{graphicx}
\usepackage{wrapfig}
\usepackage{subcaption}
\usepackage[hyperfootnotes=false]{hyperref}
\usepackage[automake,abbreviations,symbols,nogroupskip,nonumberlist]{glossaries-extra}

\newlength\Li \newlength\Lii 
\newcommand{\ignore}[1]{}

\theoremstyle{definition}

\theoremstyle{remark}

\makeatletter
\newcommand*{\rom}[1]{\expandafter\@slowromancap\romannumeral #1@}
\makeatother

\newtheoremstyle{defnopunct}
{3pt}
{3pt}
{}
{}
{\bfseries}
{ }
{.5em}
{}

\makeglossaries
\usepackage{etoolbox}
\let\bbordermatrix\bordermatrix
\patchcmd{\bbordermatrix}{8.75}{4.75}{}{}
\patchcmd{\bbordermatrix}{\left(}{\left[}{}{}
\patchcmd{\bbordermatrix}{\right)}{\right]}{}{}

\makeatletter
\patchcmd{\subsection}{\bfseries}{\itshape}{}{}
\patchcmd{\@sect}{\@addpunct.}{}{}{}
\makeatother

\renewcommand*{\glossarysection}[2][]{}

\newacronym{CDR}{CDR}{Call Data Record}
\newacronym{GPS}{GPS}{Global Positioning System}
\newacronym{OD}{OD}{Origin-Destination}
\newacronym{PEP}{PEP}{Privacy Enhanced Person}
\newacronym{GH3}{GH3}{Geohash level 3}
\newacronym{GH5}{GH5}{Geohash level 5}
\newacronym{h37}{h37}{H3 hierarchical geospatial index level 7}
\newacronym{LMIC}{LMIC}{Low and Middle Income Countries}
\newacronym{RTO}{RTO}{Return-To-Origin}
\newacronym{MSE}{MSE}{Mean Squared Error}
\newacronym{RMSE}{RMSE}{Root Mean Squared Error}
\newacronym{TDA}{TDA}{Topological Data Analysis}
\newacronym{SDG}{SDG}{Sustainable Development Goal}
\newacronym{H3}{H3}{Hierarchical Hexagonal Geospatial Indexing System}
\newacronym{GUP}{GUP}{Globally Unconnected OD Pair}

\makeatletter
\renewcommand\listoffigures{%
        \@starttoc{lof}%
}
\makeatother

\author{Alisha Foster$^\dag$, David A. Meyer$^\ddag$, Asif Shakeel$^\ddag$}
\address{$^\dag$Department of Applied Mathematics, University of Washington, Seattle, WA 98195-3925, USA}
\address{$^\ddag$Department of Mathematics, University of California, San Diego, La Jolla, CA 92093-0112, USA}
\email{alfost@uw.edu, dmeyer@ucsd.edu, ashakeel@ucsd.edu}

\title [Network-Level Measures of Mobility from Aggregated OD Data]{Network-Level Measures of Mobility from Aggregated Origin-Destination Data}

\begin{document}

\begin{abstract}
We introduce a framework for defining and interpreting collective mobility measures from
spatially and temporally aggregated origin--destination  (OD) data. Rather than characterizing
individual behavior, these measures describe properties of the mobility system itself:
how network organization, spatial structure, and routing constraints shape and channel
population movement. In this view, aggregate mobility flows reveal aspects of connectivity,
functional organization, and large-scale daily activity patterns encoded in the underlying
transport and spatial network.

To support interpretation and provide a controlled reference for the proposed
time-elapsed calculations, we first employ an independent, network-driven
synthetic data generator  in which trajectories
arise from prescribed system structure rather than observed data. This controlled setting
provides a concrete reference for understanding how the proposed measures reflect network
organization and flow constraints.

We then apply the measures to fully anonymized data from the NetMob 2024 Data Challenge,
examining their behavior under realistic limitations of spatial and temporal aggregation.
While such data constraints restrict dynamical resolution, the resulting metrics still
exhibit interpretable large-scale structure and temporal variation at the city scale.

\end{abstract}


\maketitle

\section{Introduction} \label{sec:intro}

Human mobility has become a central object of study across the physical and social
sciences, driven by the widespread availability of digitally mediated traces of movement
from mobile devices, applications, and networked services. Early work in this area relied
largely on individual-level records, such as call detail records (\glspl{CDR}), \gls{GPS} traces,
and application usage logs, which enable the reconstruction of trajectories and the
characterization of personal movement patterns. However, growing concerns over privacy and
re-identification risk~\cite{dgbcd:opcumpd,xtlzfj:trfa} have increasingly constrained access
to such fine-grained data. As a result, much contemporary mobility data is released only in
aggregated form, reporting collective \gls{OD} flows between coarse spatial
units over fixed time intervals~\cite{bbccd:amdchfc,tizzoni2014od}.

In these datasets, the study region is partitioned into a spatial grid and the observation
period into a sequence of temporal intervals. For each interval, the data typically consist
of population counts within grid cells and aggregated trip counts between origin and
destination cells, sometimes accompanied by summary statistics such as mean travel distance
or time. This form of aggregation deliberately suppresses individual identities and temporal
continuity, revealing mobility only at the collective level. While such representations are
well suited to privacy-preserving analysis, they fundamentally alter the questions that can
be asked: individual trajectories are no longer observable objects, and mobility must be
understood through the structure and evolution of flows on a network of locations. Even when auxiliary information is available, reconstructing or matching individual
mobility traces across large-scale datasets has been shown to be fundamentally limited under
aggregation, underscoring the loss of identifiability inherent in privacy-preserving mobility
representations~\cite{khdr:tmumtlsd}.
At this
level of abstraction, mobility is naturally described as flow on a spatial network, where
structure and dynamics are intertwined, rather than as collections of individual
trajectories~\cite{barthelemy2011spatialnetworks}. Viewed in this way, aggregated data provide a system-level description
of cities as organized, evolving spatial networks, aligning with broader
conceptualizations of urban structure as emergent from flows rather than static
land use~\cite{batty2013newscience}.

A substantial body of work has developed models and measures of mobility from individual
trajectory data, including displacement statistics, jump-length distributions, radius of
gyration, and related long-term indicators~\cite{ghb:uihmp,bbgj:hmma,bhg:slht}. Network constraints imposed by urban street topology have also been shown to
strongly influence observed displacement statistics, even when individual
movement rules are simple~\cite{jyz:chmlsn}.
These measures
have proven effective in characterizing regularities of personal movement and their
predictability. In contrast, the present work is not concerned with reconstructing or
approximating individual behavior. Instead, we ask what can be learned about mobility at a
macroscopic level—about network organization, connectivity, and the ability of spatial
infrastructure to channel population movement—when individuals are observable only through
their aggregated contributions to OD flows. Related efforts have also explored mobility measures derived from enriched or
hybrid datasets, such as mobile phone records augmented with geographic
information systems, highlighting the sensitivity of mobility metrics to data
representation and preprocessing choices~\cite{wtded:mhmumpr}. From this perspective, the present work is closer in spirit to
diffusion and flow processes on temporal networks than to
trajectory-based models of individual mobility, with aggregation
inducing an effective Markovian description at the population level.

Our objective is therefore to develop \emph{network-level mobility measures} derived
entirely from collective OD data. These measures are inspired in spirit by individual-based
metrics but differ fundamentally in interpretation: they characterize properties of the
mobility system itself, as revealed by population flows, rather than properties of
individual movers. In this sense, individuals act as distributed probes of the underlying
spatial and temporal structure, allowing collective movement to reveal how effectively the
network supports recurring activities, daily rhythms, and longer-term
redistribution. 

To support this perspective, we introduce a pseudo Markov-chain formulation in which
time-indexed OD matrices define a sequence of stochastic flow operators acting on an
indistinguishable population. This construction provides a principled way to reason about
time-elapsed flows across multiple aggregation intervals while remaining entirely within
the collective description imposed by privacy-preserving data release. Building on this
framework, we define several mobility measures that summarize cumulative flow behavior,
indirectness of movement, and recurrence at the network level.

The measures are examined using mobility data from the NetMob 2024 Data Challenge
\cite{nm:nm2024}, provided by Cuebiq.~\footnote{Cuebiq's terms and conditions for the NetMob 2024 Data Challenge require us to refer the reader to the following statement pertaining to  the availability of data: Aggregated data was provided by Cuebiq Social Impact as part of the Netmob 2024 conference. Data is collected with the informed consent of anonymous users who have opted-in to anonymized data collection for research purposes. In order to further preserve the privacy of users, all data has been aggregated by the data provider spatially to the \gls{GH3}, \gls{GH5}, \gls{h37} and temporally to 3-hourly, daily, weekly, and monthly levels and does not include any individual-level data records.} This dataset is part of a broader recent effort to promote open, standardized, and
comparative human mobility data products, which have expanded opportunities for
population-level analysis while also emphasizing the analytical consequences of spatial and
temporal aggregation~\cite{yltlgm:ehmrwosd}.
The dataset
covers Mexico, Indonesia, India, and Colombia, and reports population counts and OD flows at
multiple spatial and temporal resolutions. In this study, we focus on 3-hourly GH5 OD data
for the year 2019. As is typical of strongly aggregated datasets, the data exhibit temporal
coarsening and sparsity effects that limit direct interpretability at short time scales.
Rather than treating these limitations as obstacles to trajectory reconstruction, we use
them to motivate mobility indicators that are robust to aggregation and meaningful at the
scale at which the data are available.

To aid interpretation and validate the proposed constructions, we additionally employ a
separate, network-driven synthetic data generator in which trajectories arise from
externally specified structural and temporal biases. These synthetic trajectories are not
intended to model observed behavior; instead, they provide controlled settings in which the
relationship between network structure, prescribed dynamics, and resulting mobility
measures can be examined explicitly. Related work on collective human mobility from aggregated data has examined flow structures, event-driven patterns, and relationships between short- and long-term movements~\cite{bbbc:uhmfampd,jwsl:chmpttua,gyzh:uichmptceos,fdgm:mngmc,mzbwt:ursltm}. 
In contrast, the focus here is not on identifying or classifying behavioral regimes, but on developing time-elapsed measures derived from a pseudo Markov-chain model that remain interpretable under strong spatial and temporal aggregation.

A related study introduces a network-driven Markov mobility generator that produces full
synthetic trajectories and validates, in controlled settings, the consistency between
trajectory-level realizations and aggregated OD representations~\cite{ms:ntmd}.
The synthetic datasets used here are generated using the publicly available code from that
work; however, the definitions, algorithms, and analyses of time-elapsed mobility measures
presented in this paper are self-contained and do not rely on trajectory-level assumptions.

Because the proposed framework is defined entirely at the level of
time-dependent Markov transition operators, all time-elapsed mobility
measures considered here are fixed by these operators up to finite-sample
effects. Comparisons with realized trajectories therefore probe only
sampling variability, rather than providing an independent form of
empirical validation. Accordingly, validation in this work focuses on
internal consistency, identifiability, and reproducibility; all
computations are fully specified and supported by publicly available
code.

This paper is organized as follows.
Section~\ref{sec:pmcm} introduces the pseudo Markov-chain framework for
spatially and temporally aggregated OD data and discusses
limitations arising from temporal aggregation that we define as \textit{mobility-aliasing}.
Section~\ref{sec:tenet} develops time-elapsed net flow measures and demonstrates
how temporal composition reveals large-scale directional structure in both
synthetic and aggregated mobility settings.
Section~\ref{sec:tedist} introduces time-elapsed OD distance
and the associated notion of effective distance, which serves as a network-level
measure of indirectness and structural inefficiency in collective movement;
this section constitutes the primary empirical focus of the paper.

We also describe return-to-origin (\gls{RTO}) distances as a natural extension of the
time-elapsed framework, obtained by specializing OD distances
to coincident endpoints.
While RTO measures provide a collective analogue of long-term mobility
indicators such as radius of gyration, we do not pursue a detailed empirical
analysis here, instead outlining their definition and interpretation as a
direction for future work.

We conclude in Section~\ref{sec:conc} with a discussion of implications,
limitations, and directions for further research.

\section{Pseudo Markov-Chain Model} \label{sec:pmcm}

To analyze collective mobility using spatially and temporally aggregated
OD data, we introduce a pseudo Markov-chain formulation.
At the level of aggregation imposed by privacy-preserving data release,
individual identities and movement histories are not observable. As a result,
mobility can only be described probabilistically in terms of transitions
between locations. While the resulting construction resembles a time-inhomogeneous Markov chain,
the Markovian structure arises from aggregation and indistinguishability rather
than from assumptions about individual behavior~\cite{n:mc}.

We adopt the term \emph{privacy-enhanced person} (\gls{PEP}) to denote an abstract
realization of this description: a notional, indistinguishable mover whose
transitions reflect observed collective flows rather than an identifiable
trajectory. PEPs should not be interpreted as approximations to real
individuals; instead, they serve as analytical carriers of population flow,
providing a convenient representation for reasoning about aggregate mobility
without assigning persistent identities.

For each time-step $t$, we construct a stochastic transition matrix $M^t$
encoding the probability of moving from an origin location to a destination
location within that interval. To extend inference beyond single time slices,
we define \emph{time-elapsed transition matrices} that capture the cumulative
effect of successive flow operators over multiple intervals. These matrices
form the basis for estimating longer-term, collective mobility quantities such
as net population flow, cumulative travel distance, and recurrence statistics,
all defined entirely at the aggregate level.

Depending on the chosen spatial and temporal resolution, observed trips may
predominantly occur within metropolitan regions or between distant regions.
Consequently, mobility structure is often localized in space and time, and it
is natural to focus on connected components at the scales of interest rather
than on entire national networks. The pseudo Markov-chain model itself is defined generically for aggregated OD
data and is consistent with standard treatments of diffusion and transport on
temporal networks, where Markovian structure arises from coarse-graining rather
than behavioral assumptions~\cite{n:mc,masuda2017temporal}.

\subsection{Network-Driven Synthetic Data Generator}
\label{subsec:synthetic}

To support interpretation and validation of the proposed time-elapsed mobility
measures, we employ a network-driven synthetic data generator that produces
controlled, aggregate OD flows.
The goal of this construction is not to reproduce empirical mobility patterns,
but to provide a reference setting in which the relationship between network
structure, prescribed dynamics, and resulting collective mobility measures can
be examined explicitly. For clarity, we first describe a controlled, network-driven synthetic mobility
setting that motivates the construction, before introducing empirical
aggregated OD data.

The generator is adapted from the network-driven Markov dynamics introduced in~\cite{ms:ntmd}, where individual mobility trajectories arise from
time-dependent stochastic operators defined on a spatial network.
In this work, we use the same framework solely to generate aggregated OD flows, without
retaining individual identities or paths, and in a form consistent with the
privacy-preserving, collective description assumed throughout this paper.

The synthetic mobility network is constructed using the H3 hierarchical
hexagonal spatial index~\cite{uberh3}, rather than the geohash-based grids
used in the NetMob dataset.
We employ H3 resolution~6 (R6), for which adjacent cell centers are separated by
approximately $6.4\,\mathrm{km}$ on average,
 providing a regular and isotropic
spatial tessellation suitable for network-based mobility modeling.
Time is discretized into $30$-minute intervals, a resolution commensurate with
typical urban travel speeds at this spatial scale.
\begin{figure}[H]
\centering
\includegraphics[width=0.4\linewidth]{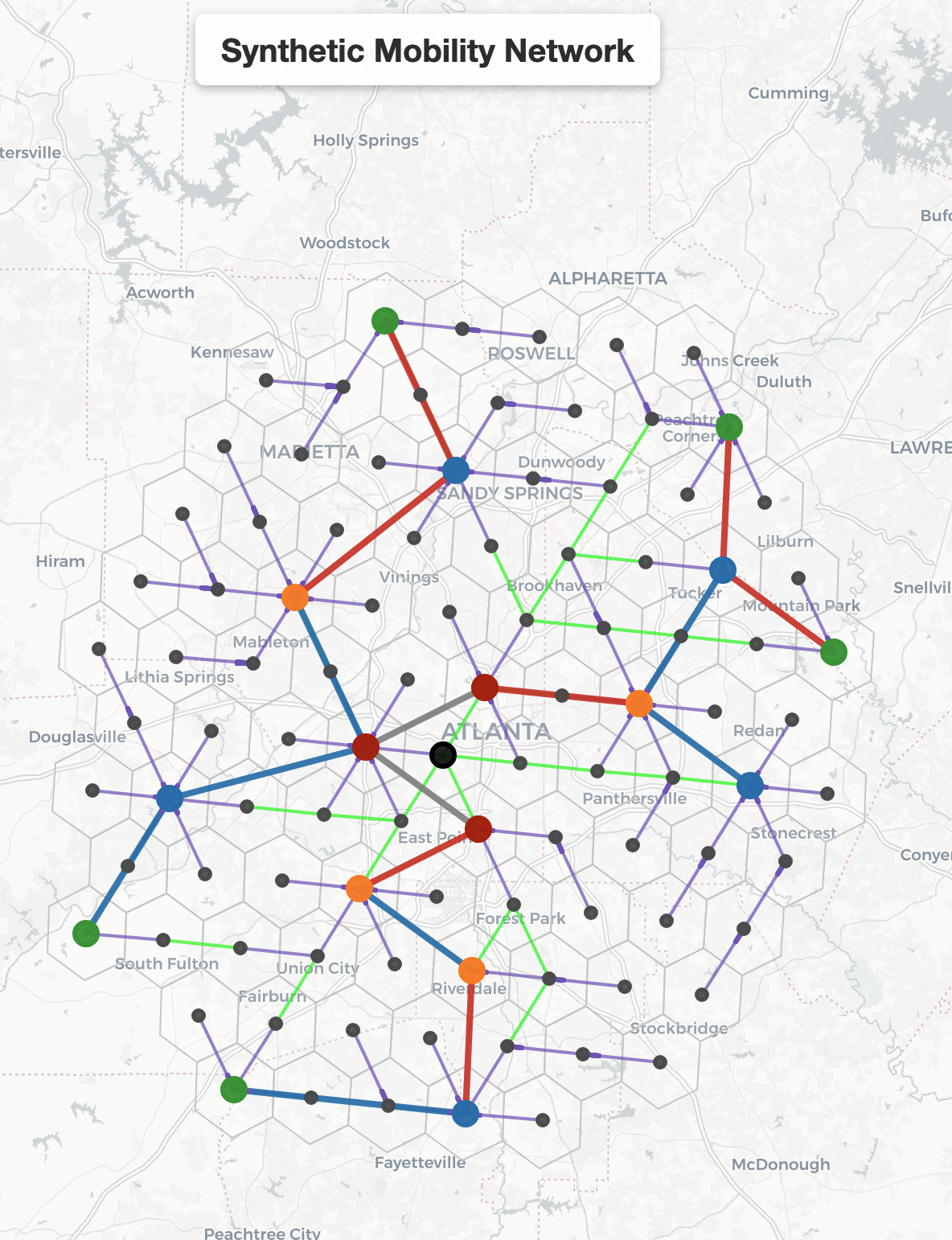}
\caption{
Synthetic mobility network used in all experiments.
Nodes correspond to \gls{H3} resolution~6 cells over the Atlanta metropolitan area.
Thicker nodes denote hub locations, with node color indicating distance bands
from the designated center.
Thicker edges correspond to high-capacity metro connections, while thinner edges
represent local connections.
This network serves as the fixed spatial structure on which time-dependent
Markov dynamics generate synthetic mobility flows.
}
\label{fig:synth-network}
\end{figure}

\noindent\emph{Interpretation of synthetic time indices:}

For clarity, the calendar dates associated with the synthetic experiments
(e.g., 2025--06--01 to 2025--06--29) carry no intrinsic semantic meaning.
They are used solely to index a sequence of consecutive time steps and to
define a finite range of days over which periodic dynamics are observed.
The synthetic data generator does not model real calendar effects,
seasonality, or external events, and the choice of dates is arbitrary.
All results derived from the synthetic data depend only on the relative
ordering and duration of time steps, not on their absolute placement on
the calendar.

The network is instantiated over real geographic locations using the Atlanta
metropolitan area as a spatial backdrop.
Nodes correspond to H3~R6 cells, while edges encode admissible movements defined
by a prescribed corridor-and-feeder structure, including hub nodes,
high-capacity metro edges between selected hubs, and lower-capacity local
connections.
Figure~\ref{fig:synth-network} shows the resulting network, which is fixed
across all synthetic experiments.
Full construction details are given in~\cite{ms:ntmd}.

To induce interpretable large-scale movement patterns, the dynamics incorporate
a deliberately simplified center--periphery structure.
A small set of cells near a geographic and network-defined reference point is
designated as the \emph{center}, while all remaining cells are treated as
peripheral.
This notion of center is independent of network degree: hub nodes serve a
distinct structural role by enabling aggregation and long-range connectivity,
and may but need not coincide with central cells.

Mobility is generated by a prescribed, time-varying, gravity-like model defined
on a fixed spatial network.
At each time-step, locations are assigned activity levels and stay
probabilities, and movements are biased toward nearby, higher-activity nodes,
with additional structural emphasis placed on hubs and high-capacity corridors.
A simple center--periphery potential, defined by graph distance from a selected
set of reference cells, provides a consistent notion of inward and outward
movement along the network.
During morning-like phases, this gravity field preferentially attracts movement
toward central locations, while during afternoon and evening phases the bias is
partially reversed, favoring outward redistribution.
Additional mechanisms—such as elevated stay probabilities at hubs and
time-dependent modulation of long-range metro links—promote accumulation,
through-flow, and delayed release of population.

The synthetic mobility network is constructed to be strongly connected as a
static graph.
In addition, the time-dependent mobility dynamics are designed so that the
product of transition matrices over one 24-hour cycle is primitive.
This guarantees the existence and uniqueness of a periodic fixed point for the
time-dependent Markov dynamics~\cite{ms:ntmd}.
Because the prescribed transition schedule is strictly periodic with a
24-hour cycle, the corresponding time-dependent Markov dynamics admit a
periodic fixed point for each of the daily cyclic products starting at any time-step of the day.
This fixed point we choose is the one that corresponds to the population distribution at the start
of each daily cycle (midnight), expressed as a probability distribution
over all spatial tiles in the network.
In all synthetic experiments reported here, simulations are initialized
at this periodic fixed point, so that observed OD trip-counts for each time-step and ensuing flow calculations reflect the
steady cycling behavior of the system rather than transient relaxation.
The existence and computation of such periodic fixed points for
network-driven Markov mobility models are established in~\cite{ms:ntmd}.

\begin{figure}[H]
\centering
\includegraphics[width=0.75\linewidth]{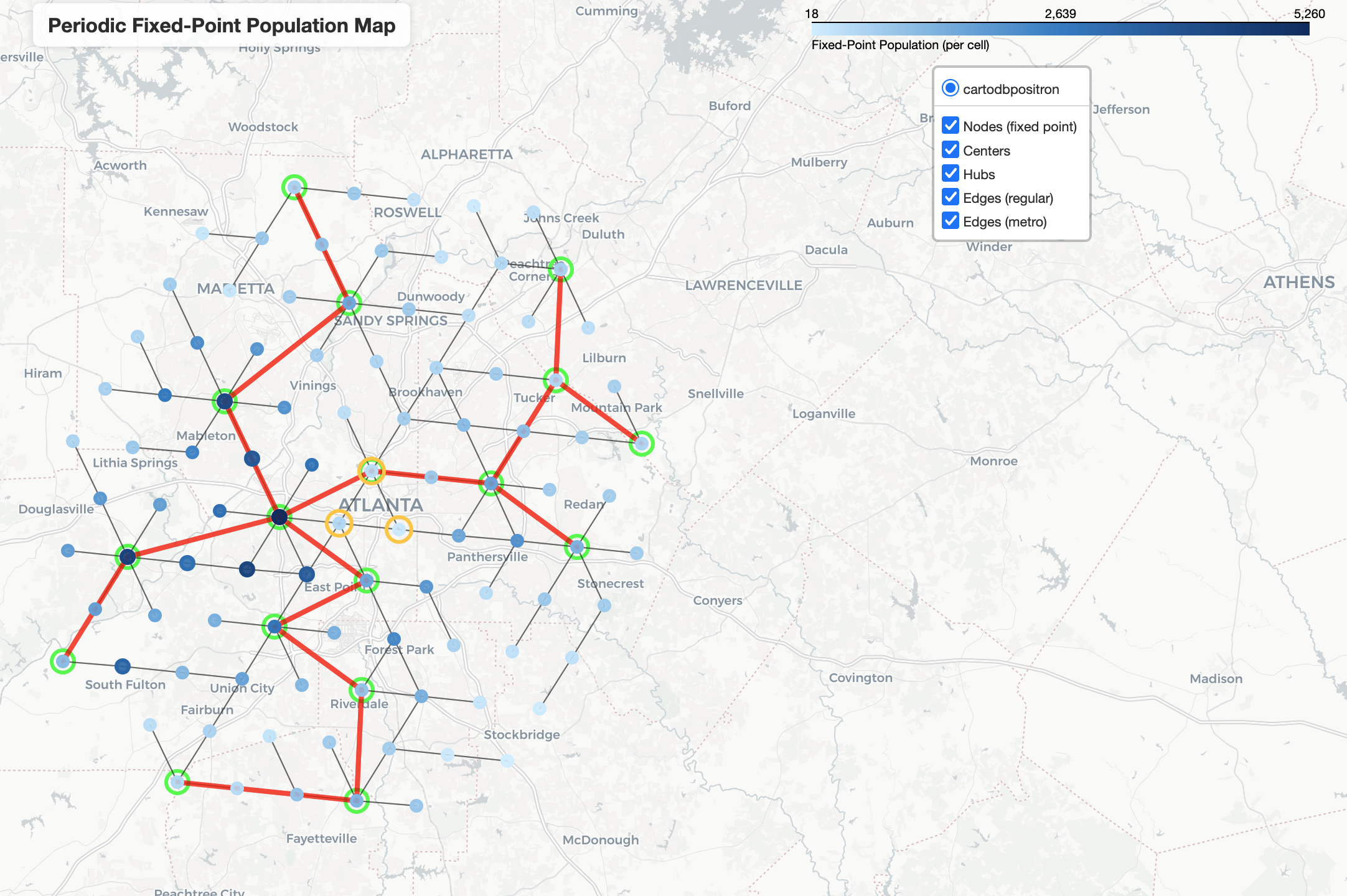}
\caption{
Periodic fixed-point population distribution of the synthetic mobility model,
corresponding to the midnight distribution of population-equivalent persons
across H3 tiles.
Darker colors indicate higher population density.
Amber-circled nodes denote the designated center of the network, while all
remaining nodes form the periphery.
}
\label{fig:synth-fixed-point}
\end{figure}

Synthetic data is realized by simulating $120{,}000$ population-equivalent
persons (PEPs) evolving under these dynamics.
Resulting trajectories are aggregated into OD matrices at each time-step and
over time-elapsed windows. In this paper we work exclusively with the aggregated synthetic OD data; statistical agreement with trajectory-level realizations is established in~\cite{ms:ntmd}.
All aggregation and visualization procedures are implemented in open-source
code.

Because the network structure and temporal biases are prescribed rather than
inferred, the resulting OD flows admit direct interpretation.
In the sections that follow, we therefore use this synthetic generator to
establish reference behavior for time-elapsed net flows, before applying the
same constructions to strongly aggregated real-world mobility data.

\subsection{Aggregated OD Data}
\label{subsec:gendatadesc}

Having introduced the framework and its behavior in a controlled synthetic
setting, we now apply the pseudo Markov-chain model to time-indexed,
spatially aggregated OD data of the type commonly
released in privacy-preserving mobility studies.
 Such data describe population movement
between discrete spatial locations over fixed temporal intervals and arise
both from empirical measurement pipelines and from network-driven synthetic
mobility generators.

In this paper, we apply the framework to the NetMob 2024 Data Challenge dataset
\cite{nm:nm2024,zpgm:nmd}, provided by Cuebiq. The dataset reports aggregated OD
flows and population counts for several countries, including Mexico,
Indonesia, India, and Colombia, at multiple spatial and temporal resolutions.
All data are fully anonymized prior to release and contain no individual-level
records.

In the NetMob dataset, time is discretized into fixed intervals (3-hourly,
daily, weekly, or monthly), and space is discretized into grid cells defined by
geohash-based centroids. Spatial resolutions include \gls{GH5} (square grid, side
length approximately 4.9 km), \gls{GH3} (square grid, side length approximately
156.5 km), and H3 resolution~7 (h37; hexagonal grid, side length approximately 1.41 km). All
computations in this paper use the 3-hourly GH5 OD data for the year 2019.

Each trip record is indexed by a triple consisting of a time-step, an origin
cell, and a destination cell. For each OD pair, the data
include the aggregated trip count as well as summary statistics such as the
mean, median, and standard deviation of trip distance and trip duration. OD
pairs with trip counts below a threshold of 10 are omitted. As a result, the
data do not guarantee that trips observed within the same or consecutive
time-steps correspond to distinct individuals. In particular, the same
individual may contribute to multiple OD pairs within a single interval or
across successive intervals.

For each time-step, the OD data can be represented as a weighted directed
graph, with nodes corresponding to locations and edge weights given by trip
counts. To extend inference beyond individual time slices and to define
meaningful longer-term mobility measures, we organize these graphs into a
pseudo Markov-chain structure.

For clarity, in all subsequent comparisons we use the term
\emph{aggregated data} to refer specifically to the NetMob 2024 OD data provided
by Cuebiq, in contrast to the \emph{synthetic data} generated by the
network-driven model described in Section~\ref{subsec:synthetic}.

\subsection{Model Construction}
\label{subsec:pmcconstr}

We now formalize the pseudo Markov-chain construction for generic, time-indexed
OD flow data. The formulation is independent of how the
OD matrices are obtained and applies equally to empirically measured or
synthetically generated aggregate flows.

Consider a sequence of OD flow matrices indexed by discrete time-steps
$t = 1, \ldots, T$. Let $N$ denote the total number of spatial locations. At each
time-step $t$, the OD data define a weighted directed graph on these locations,
with edge weights representing aggregated flow counts.

Let $f^t_{ij}$ denote the observed flow from origin location $j$ to destination
location $i$ during time-step $t$. For each origin $j$, we define a
time-dependent transition probability to destination $i$ by normalizing the
outgoing flows,
\begin{equation}
m^t_{ij} = \frac{f^t_{ij}}{\sum_{i=1}^{N} f^t_{ij}},
\end{equation}
whenever the denominator is nonzero. The resulting matrix
\begin{equation}
M^t = [m^t_{ij}]
\label{eq:Mdef}
\end{equation}
is column-stochastic and represents a one-step transition operator acting on an
indistinguishable population at time-step $t$.

Mobility flows often concentrate within subsets of locations over the time
scales of interest. We therefore restrict attention to strongly connected
components of the directed graph obtained by aggregating edges over multiple
time-steps. Let $E^t$ denote the set of directed edges observed at time-step
$t$, and define
\[
E_T = \bigcup_{t=1}^{T} E^t.
\]
Let $C$ be a strongly connected component of the graph defined by $E_T$, with
$N_C = |C|$ nodes. All subsequent constructions are performed on $C$, ensuring
that the resulting transition operators are irreducible and free of absorbing
states.

Assuming conditional independence of PEP transitions across time-steps given the
time-dependent transition operators inferred from aggregated flows, the time-elapsed transition matrix up to
time-step $t$ is defined by
\begin{equation}
A^t = \prod_{k=1}^{t} M^k = M^t A^{t-1}, \quad A^0 = \mathbb{I}.
\label{eq:Adef}
\end{equation}
The matrix $A^t$ encodes cumulative transition probabilities over successive
intervals and provides the foundation for time-elapsed, flow-based mobility
measures.

\subsubsection{Aggregation-Induced Limitations and Mobility-Aliasing}
\label{subsec:mobalias}

A key limitation of strongly aggregated OD data arises from the mismatch
between the temporal resolution of the data and the typical speeds of movement.
At a given aggregation interval, individuals may traverse multiple spatial
cells, while the OD data record only coarse OD transitions.

This effect is evident in the NetMob 3-hourly GH5 data. Median travel distances
and durations imply typical travel times between adjacent cells that are
substantially shorter than the aggregation interval, allowing an individual to
contribute to multiple OD pairs within a single time-step or across successive
time-steps. As a result, the same individual may be repeatedly counted as part
of distinct aggregated trips.

We refer to this phenomenon as \emph{mobility-aliasing}, by analogy with
temporal aliasing in signal analysis. Mobility-aliasing is not specific to the
NetMob dataset, but is a general feature of strongly aggregated,
privacy-preserving mobility data. Reducing it would require finer temporal
aggregation or post-processing techniques to infer sub-interval transitions.
We outline a possible algorithmic compensation strategy in
Appendix~\ref{appdx:mobaliasalg}. In the present work, we proceed using the
original aggregated data. While this limits interpretability at very short time
scales, it remains sufficient for illustrating the pseudo Markov-chain
framework and for defining collective, time-elapsed mobility measures at the
level of aggregation imposed by privacy-preserving data release.

\section{Time-elapsed OD net flows}
\label{sec:tenet}

In this section, we build on the pseudo Markov-chain model to compute
\emph{time-elapsed net flows} between locations.
Starting from the total outgoing population at each origin at the beginning
of a time-elapsed interval, we estimate the \emph{net} number of trips between
each OD pair over that interval.
The resulting quantities summarize directional imbalances in collective
movement and provide a natural, flow-based description of large-scale mobility
structure (readers primarily interested in interpretation may skip
the derivation and proceed directly to the figures and discussion in
Sections~\ref{subsec:tenet-synth} and~\ref{subsec:tenetres}).

\subsection{Derivation}
\label{subsec:tenetcalc}

For each ordered OD pair $(i,j)$, we define the antisymmetric movement
function over edges $(k,l)$ in a strongly connected component $C$ at
time-step $t$ by
\begin{equation*}
s^t_{kl}(i,j)
=
\delta_i(k)\delta_j(l)
-
\delta_i(l)\delta_j(k),
\end{equation*}
where $\delta$ denotes the Kronecker delta.
This function takes value $+1$ at $(i,j)$, $-1$ at $(j,i)$, and zero elsewhere.

Let the set of ordered edges be
\[
E^o=\{(i,j)\mid i<j,\ i,j\in C\}.
\]
The probability that a generic individual is at location $j$ at the beginning
of the first time-step is estimated as
\begin{equation}
p^1_j=\frac{\sum_i f^1_{ij}}{\sum_{i,j} f^1_{ij}},
\label{eq:genprob}
\end{equation}
and the probability that this individual transitions from $j$ to $i$ over the
time-elapsed interval up to time-step $t$ is
\[
p^t_{ij}=p^1_j a^t_{ij},
\]
where $A^t=[a^t_{ij}]$ is the time-elapsed transition matrix defined in
Eq.~\eqref{eq:Adef}.

Let
\[
n^1=\sum_{i,j} f^1_{ij}
\qquad\text{and}\qquad
n^1_j=\sum_i f^1_{ij}=n^1 p^1_j
\]
denote the total population and the population at location $j$ at the
beginning of the interval, respectively.
The mean time-elapsed \emph{net trip-count} over the ordered edges $E^o$ is
then given by
\begin{align*}
\bar{s}^t(i,j)
&=
\sum_{k,l} s^t_{kl}(i,j)\, n^1 p^t_{kl} \\
&=
n^1(p^t_{ij}-p^t_{ji}) \\
&=
a^t_{ij} n^1_j - a^t_{ji} n^1_i.
\end{align*}
Positive values of $\bar{s}^t(i,j)$ indicate a net flow from $j$ to $i$,
while negative values indicate net flow in the opposite direction.

\subsection{Synthetic validation and interpretation}
\label{subsec:tenet-synth}

We first evaluate time-elapsed net flows using the network-driven synthetic
mobility generator introduced in Section~\ref{subsec:synthetic}.
In this controlled setting, both the spatial network and the time-dependent
transition biases are specified explicitly, allowing the resulting flow
patterns to be interpreted directly in terms of known structural and dynamical
features.

\begin{figure}[H]
\centering
\begin{subfigure}{.45\textwidth}
\centering
\includegraphics[width=\linewidth]{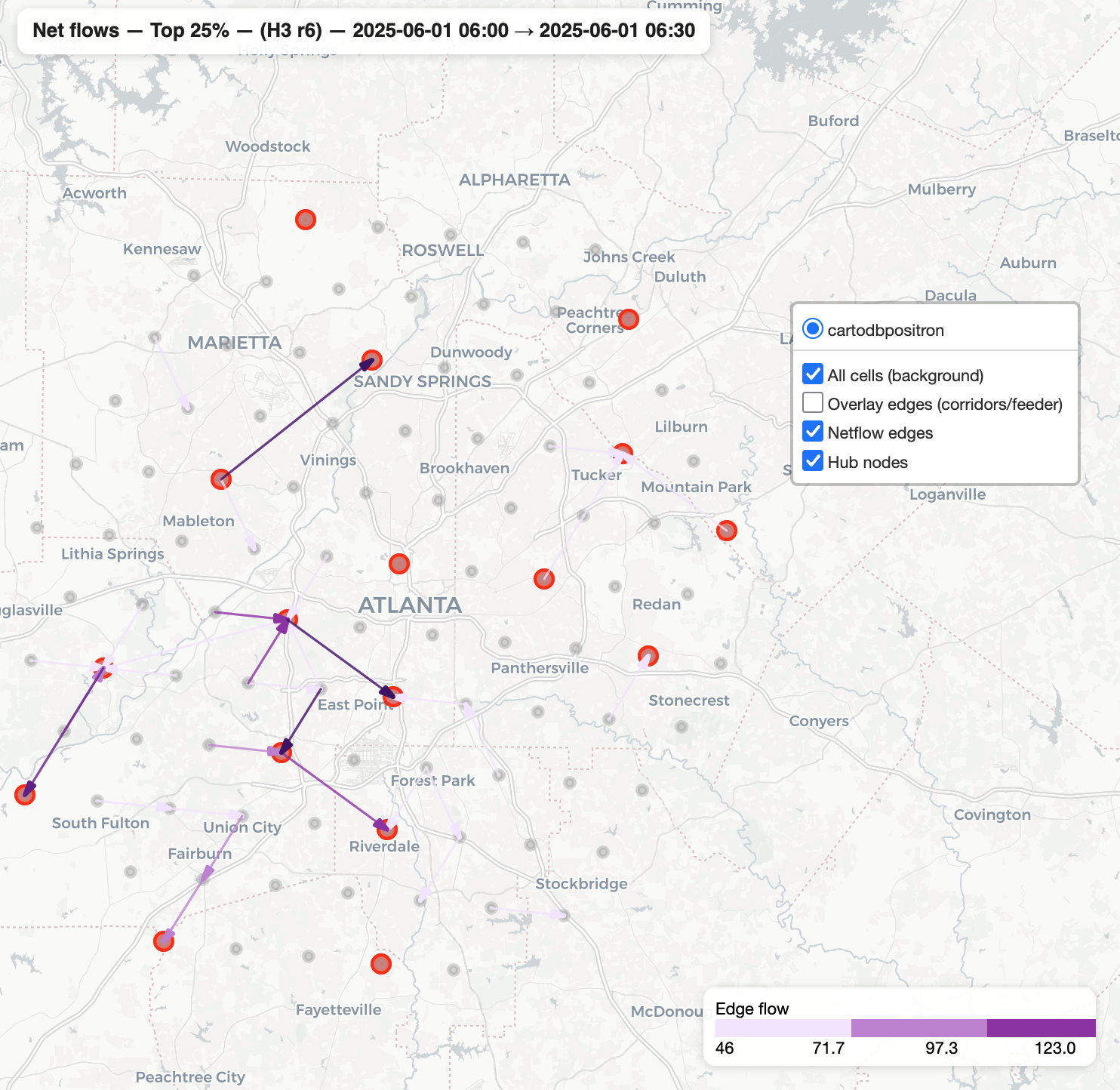}
\caption{06:00--06:30}
\label{fig:synth-am-single}
\end{subfigure}\hfill
\begin{subfigure}{.45\textwidth}
\centering
\includegraphics[width=\linewidth]{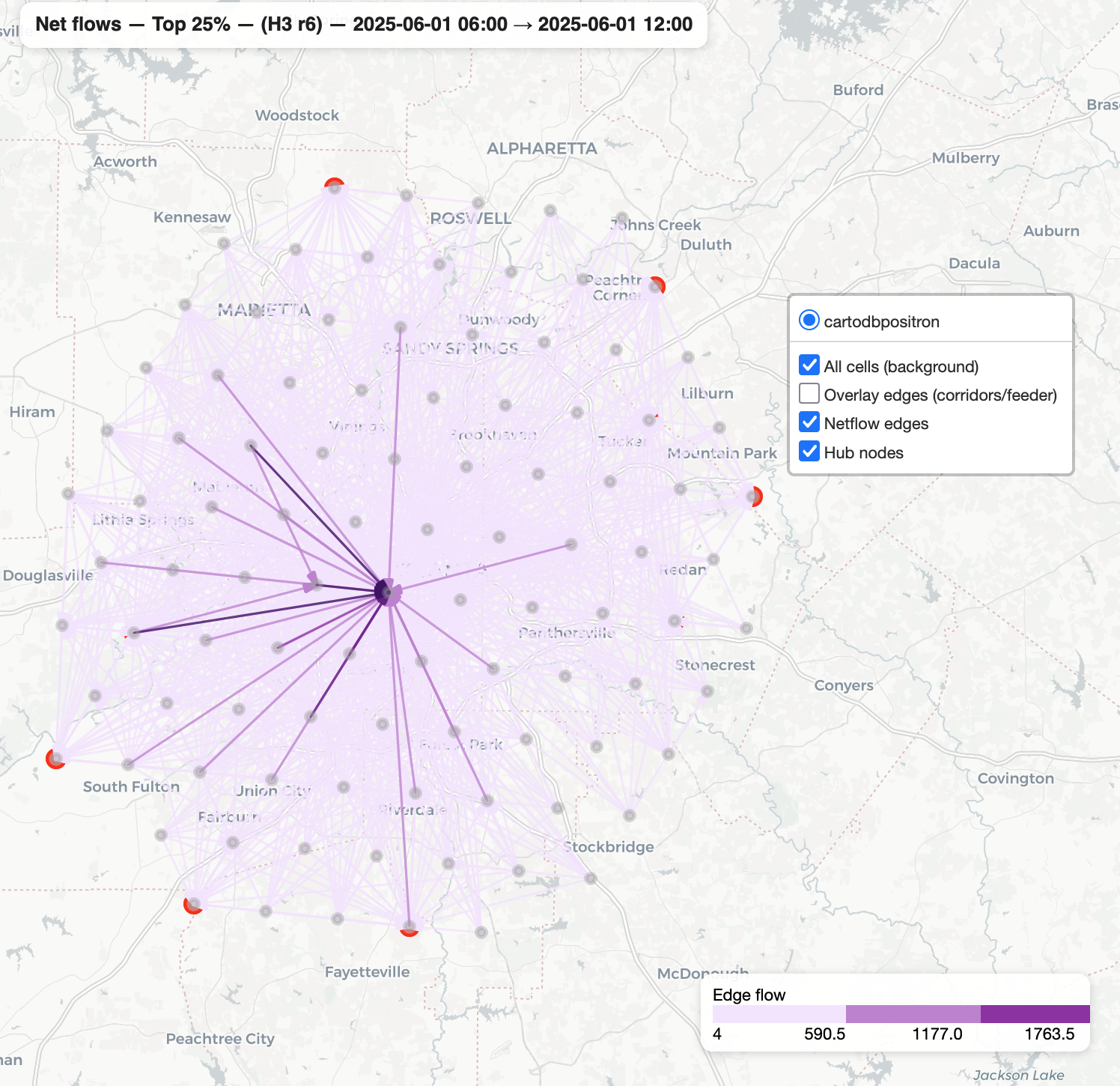}
\caption{06:00--12:00}
\label{fig:synth-am-te}
\end{subfigure}
\caption{
Synthetic morning net flows on an H3 resolution~6 spatial grid.
Top $75^{\text{th}}$ percentile of net flows shown.
Extending the time-elapsed window transforms sparse, local flows into a
coherent radial pattern directed toward central hub nodes, revealing
network-scale inbound structure.
}
\end{figure}

Figures~\ref{fig:synth-am-single} and~\ref{fig:synth-am-te} compare morning net
flows computed over a single 30-minute interval (06:00--06:30) and a longer
time-elapsed window (06:00--12:00), showing the top $75^{\text{th}}$ percentile of net flows.
At the single-interval scale, net flows are sparse and localized, with only a
small number of edges exhibiting strong directional imbalance.
These flows tend to connect nearby locations and do not yet reveal a clear
network-level organization.

When the time-elapsed window is extended, a markedly different structure
emerges.
Dominant net flows organize into a coherent radial pattern directed from
peripheral locations toward a small number of central hub nodes.
Longer-range edges enter the top $75^{\text{th}}$ percentile, reflecting sustained inflow toward
the urban core across successive intervals.
This extension sharpens directional
structure rather than merely increasing density, as short-range fluctuations
are suppressed through temporal composition.

\begin{figure}[H]
\centering
\begin{subfigure}{.45\textwidth}
\centering
\includegraphics[width=\linewidth]{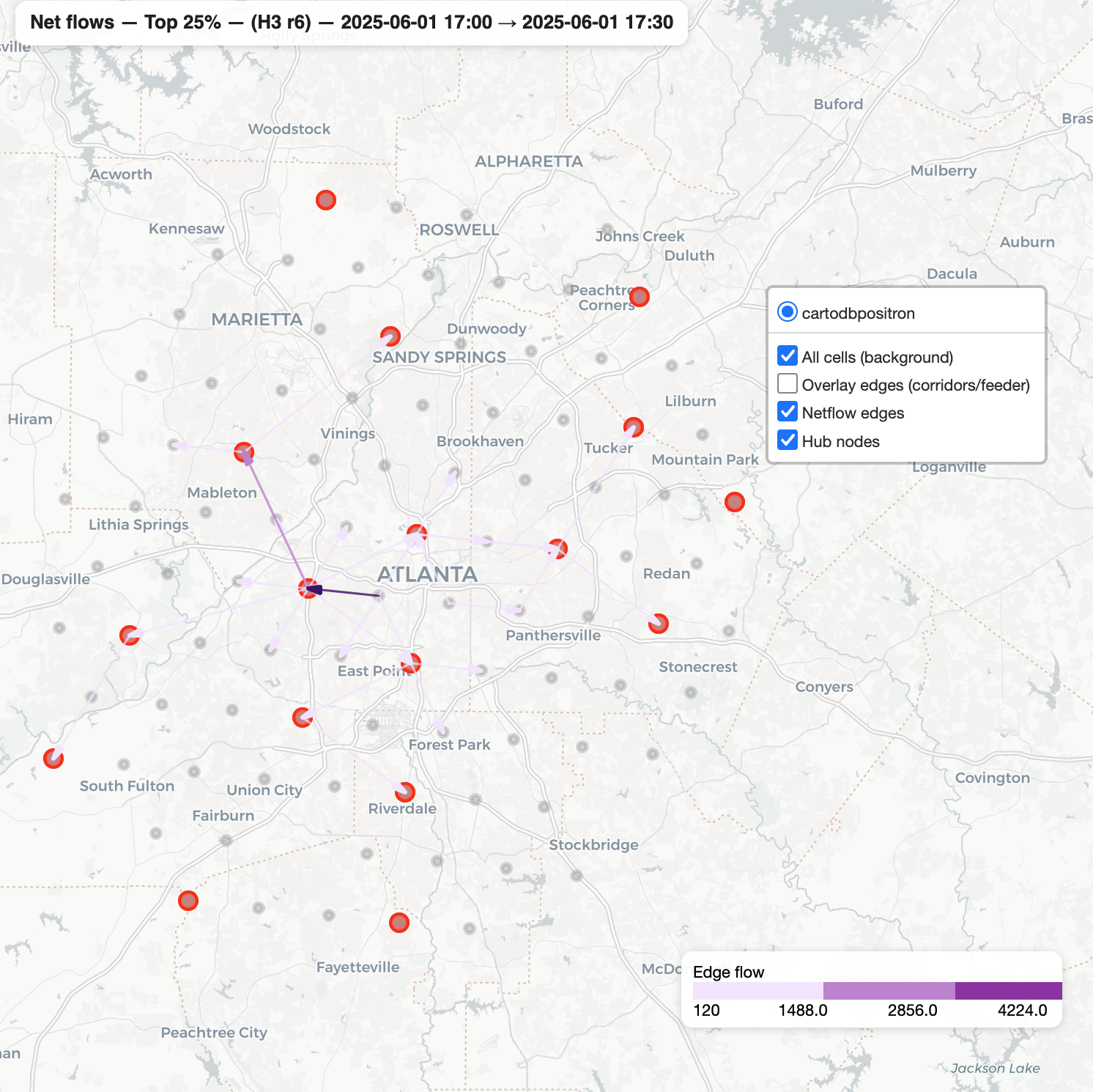}
\caption{17:00--17:30}
\label{fig:synth-pm-single}
\end{subfigure}\hfill
\begin{subfigure}{.45\textwidth}
\centering
\includegraphics[width=\linewidth]{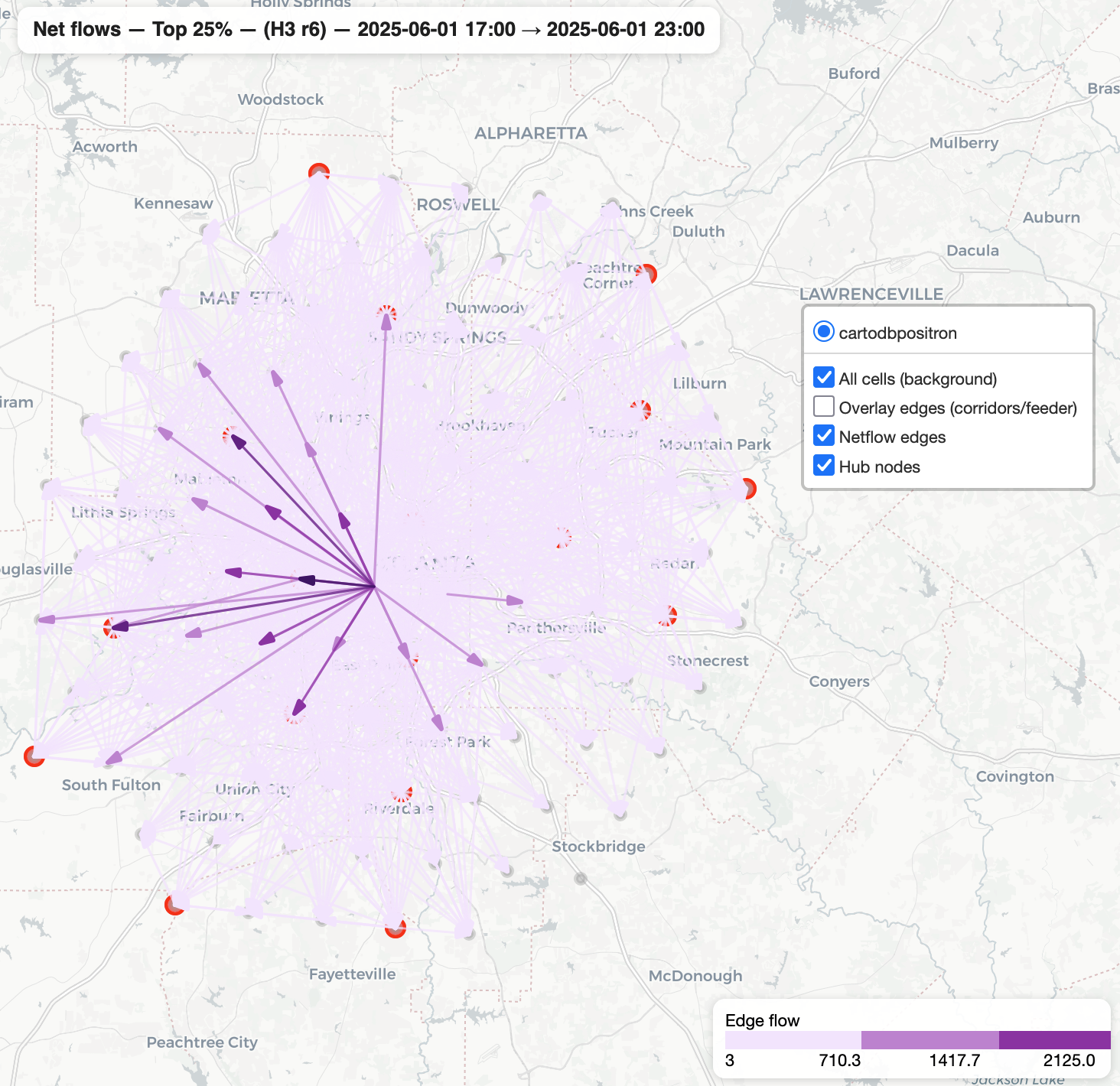}
\caption{17:00--23:00}
\label{fig:synth-pm-te}
\end{subfigure}
\caption{
Synthetic afternoon and evening net flows on an H3 resolution~6 grid.
Top $75^{\text{th}}$ percentile of net flows  shown.
Extending the time-elapsed window reveals a clear reversal of dominant flow
directions, with outward redistribution from central hubs toward peripheral
locations.
}
\end{figure}

Figures~\ref{fig:synth-pm-single} and~\ref{fig:synth-pm-te} show the
corresponding afternoon and evening net flows for the intervals
17:00--17:30 and 17:00--23:00.
In the single-interval map, net flows are again weak and fragmented, with no
dominant global direction.
Edges connect nearby locations, and outward movement from central hubs is only
partially visible.

Over the extended time-elapsed window, this pattern reorganizes into a clear
outward redistribution from central hub nodes toward the periphery.
Compared to the morning period, the flow field is more diffuse, with multiple
outgoing directions and a broader spatial spread.
Nonetheless, the dominant reversal relative to the morning inflow is clearly
resolved, with hub nodes acting as net sources rather than sinks.

This contrast between morning and afternoon phases is particularly sharp in
the synthetic setting, reflecting the finer temporal resolution and the
absence of long-distance hops within individual time-steps.

Overall, the synthetic results show that, when temporal resolution is
commensurate with spatial scale and transitions occur primarily between
adjacent locations, time-elapsed net flows can transform weak, local
imbalances into coherent, network-scale movement patterns through temporal
composition.
In this setting, extending the time window sharpens directional organization
by suppressing short-term fluctuations arising from stochastic path exploration
and reinforcing persistent structural biases encoded in the transition dynamics.
These controlled results establish reference behavior for interpreting how
time-elapsed net flows behave under less favorable conditions in aggregated
mobility data.

\subsection{Results from aggregated data}
\label{subsec:tenetres}

Unlike the synthetic setting, the NetMob data are reported at a coarse temporal
resolution of 3 hours, which permits long-distance transitions within a single
time-step and weakens the correspondence between temporal aggregation and
spatial scale.
As a result, short-time dynamics are less directly interpretable, and the
behavior of time-elapsed net flows must be assessed under conditions where
individual transitions may already span large portions of the metropolitan
area.

\begin{figure}[H]
\centering
\begin{subfigure}{.45\textwidth}
\centering
\includegraphics[width=\linewidth]{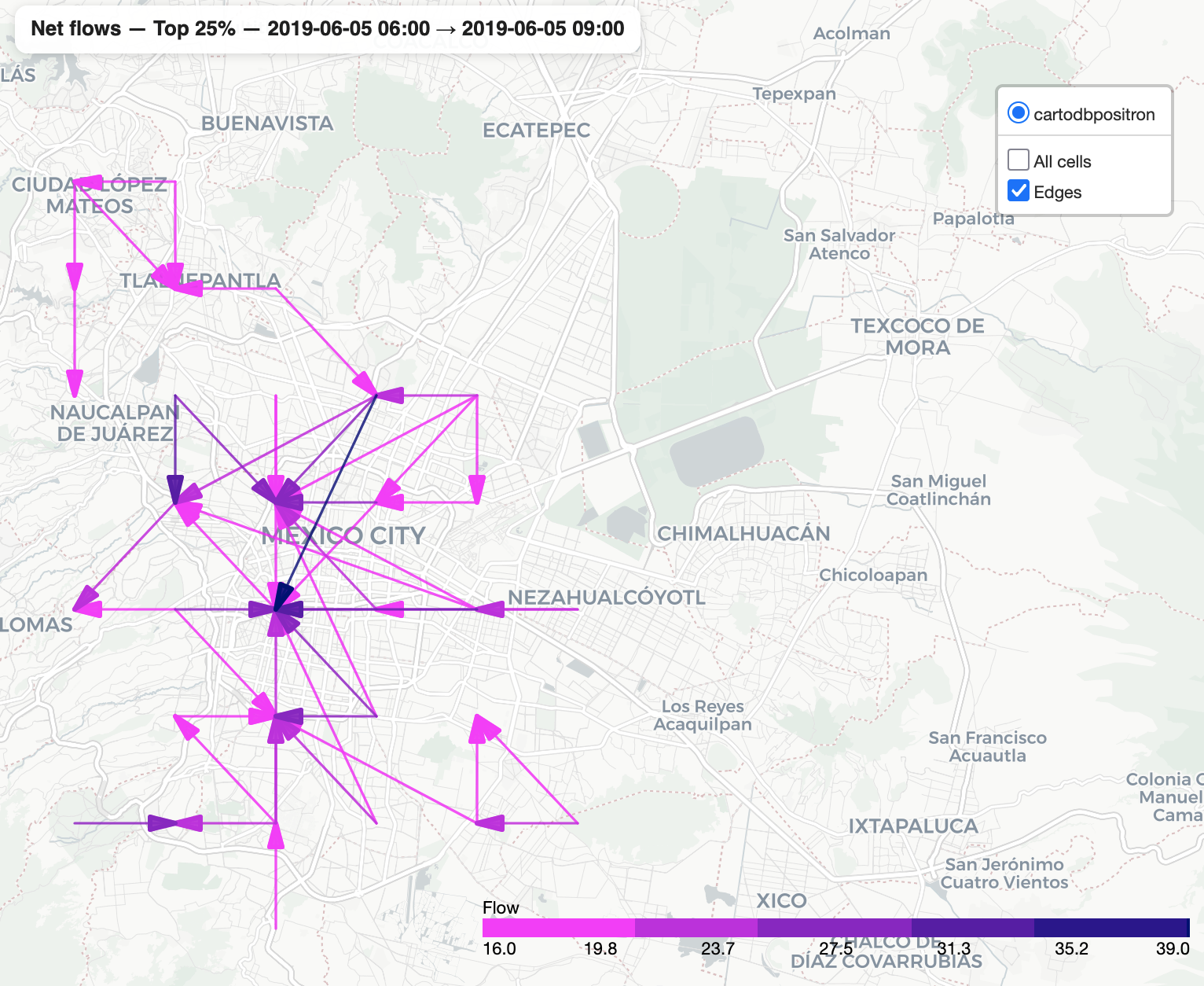}
\caption{06:00--09:00}
\label{fig:mx-am-single}
\end{subfigure}\hfill
\begin{subfigure}{.45\textwidth}
\centering
\includegraphics[width=\linewidth]{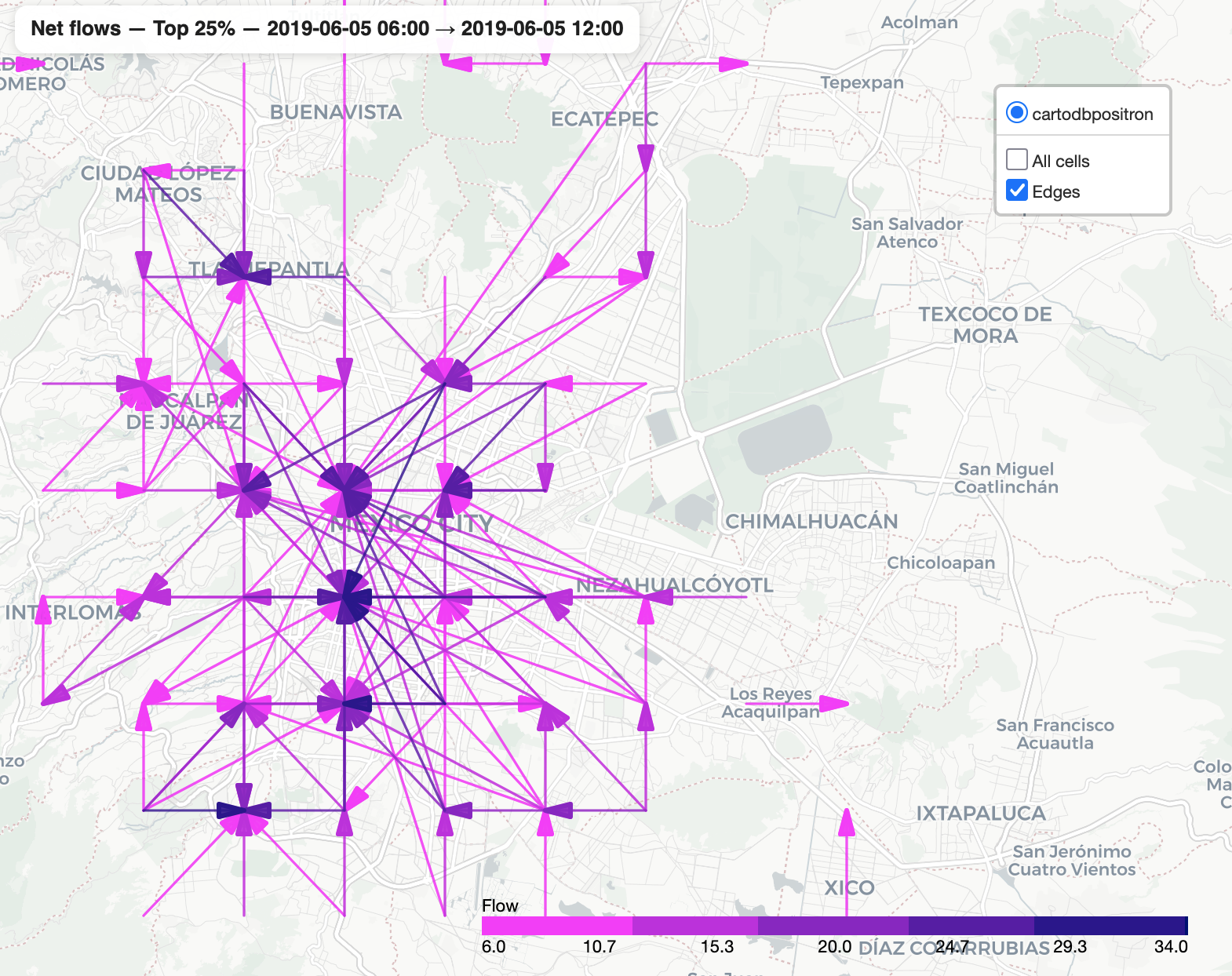}
\caption{06:00--12:00}
\label{fig:mx-am-te}
\end{subfigure}
\caption{
Mexico City, 06--05--2019.
Morning time-elapsed net flows showing the top $75^{\text{th}}$ percentile of net flows.
Extending the time-elapsed window incorporates longer-distance inflows toward
central locations, increasing both the density and spatial reach of inward
flows.
}
\end{figure}

Figures~\ref{fig:mx-am-te} and~\ref{fig:mx-am-single} compare morning net flows
computed over a single 3-hour interval (06:00--09:00) and a longer time-elapsed
window (06:00--12:00), showing the top $75^{\text{th}}$ percentile of net flows.
In both cases, the dominant net flows are directed toward a small number of
central locations, producing dense, radially inward patterns centered on the
urban core.
When the time-elapsed window is extended from 06:00--09:00 to 06:00--12:00,
the set of contributing origins expands outward, with longer-distance flows
from the metropolitan periphery entering the top $75^{\text{th}}$ percentile.
This shift reflects sustained occupancy of central locations combined with
continued inflow from progressively more distant origins, consistent with
commuters remaining in the city center while additional arrivals accumulate
over time.
By contrast, in the single-interval map, the strongest net flows originate
predominantly from locations closer to the center.

\begin{figure}[H]
\centering
\begin{subfigure}{.45\textwidth}
\centering
\includegraphics[width=\linewidth]{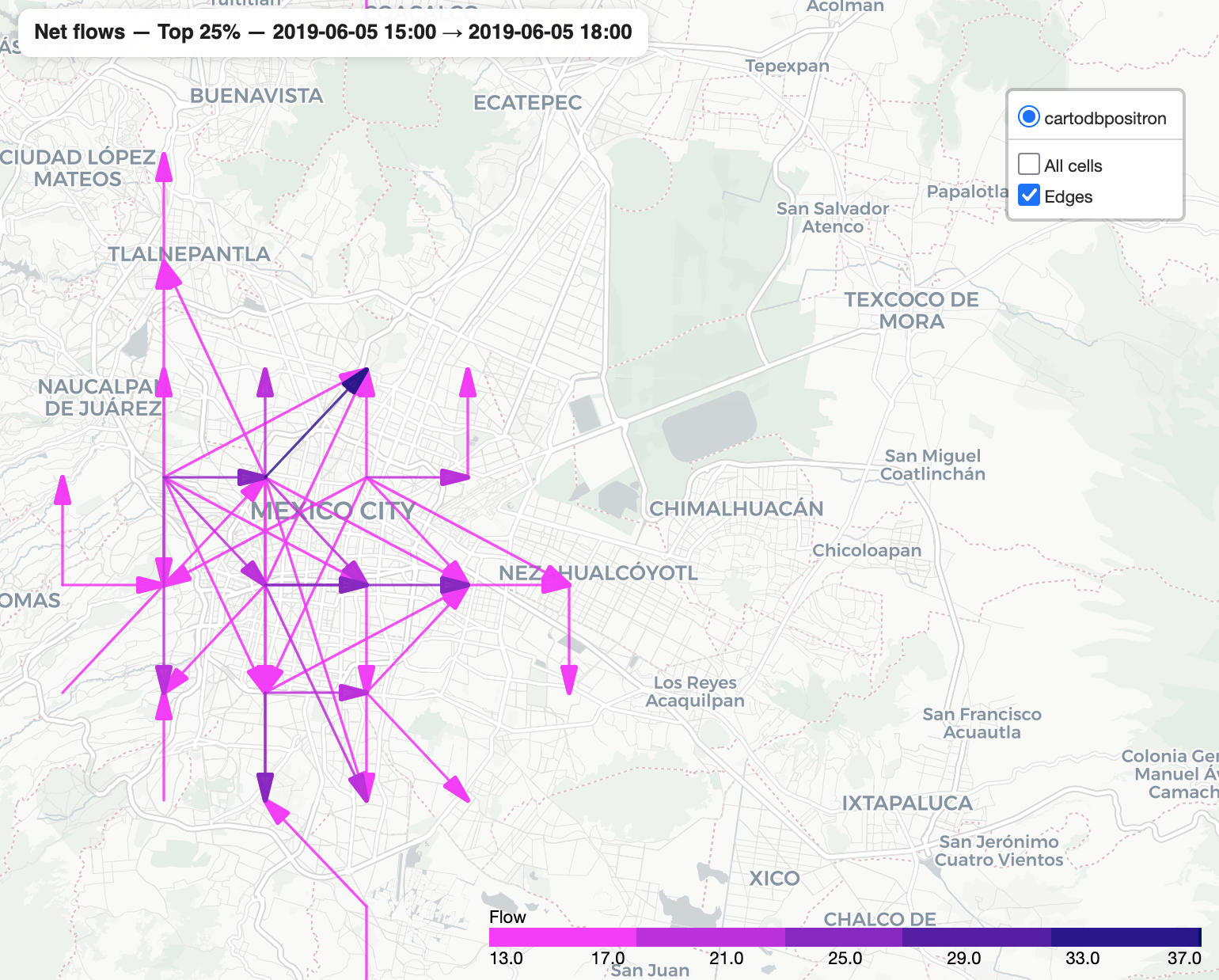}
\caption{15:00--18:00}
\label{fig:mx-pm-single}
\end{subfigure}\hfill
\begin{subfigure}{.45\textwidth}
\centering
\includegraphics[width=\linewidth]{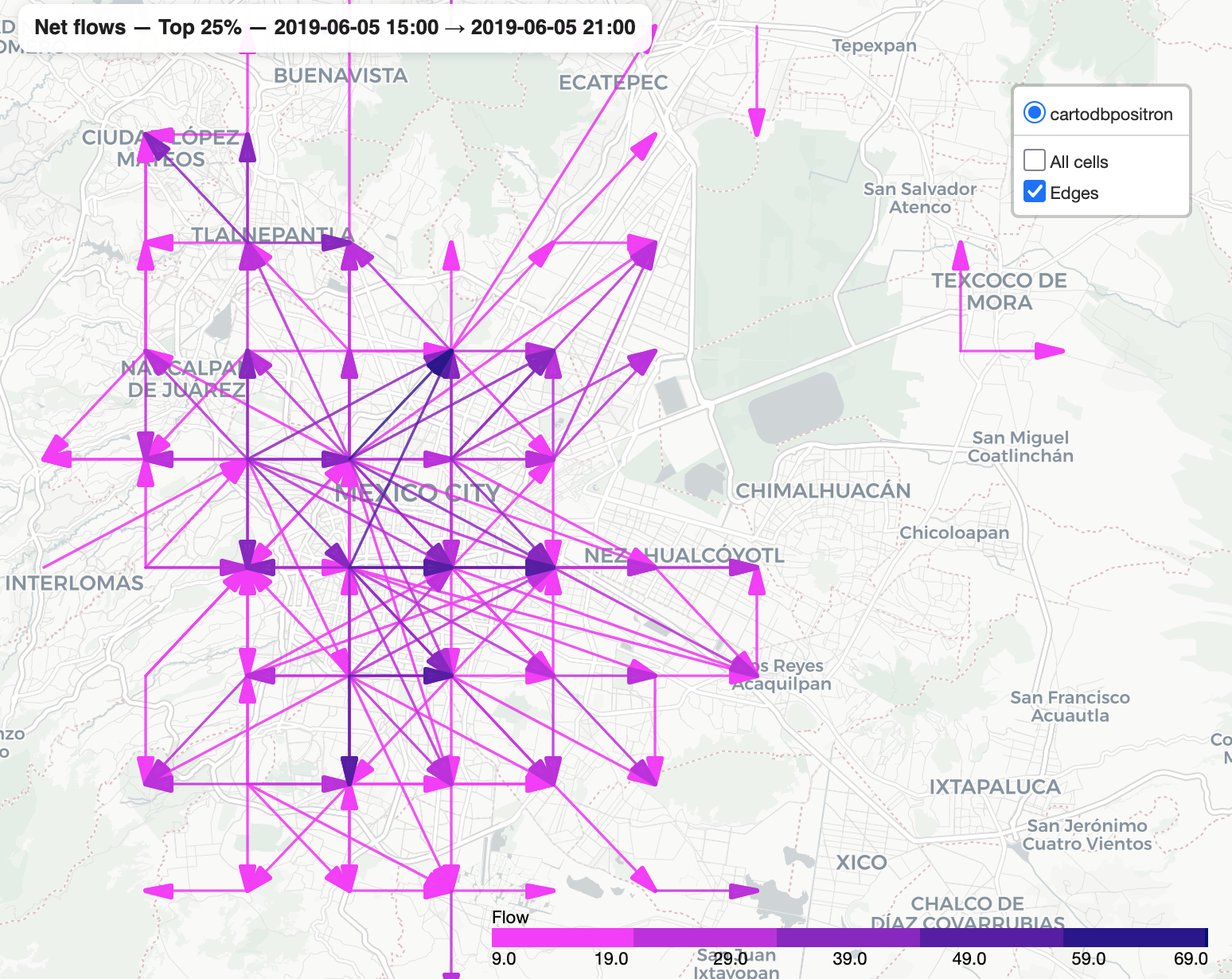}
\caption{15:00--21:00}
\label{fig:mx-pm-te}
\end{subfigure}
\caption{
Mexico City, 06--05--2019.
Afternoon and evening time-elapsed net flows showing the top $75^{\text{th}}$ percentile of net flows. 
Net flow directions partially reverse relative to the morning period, with
increased outward redistribution from central locations, though with reduced
directional coherence.
}
\end{figure}

Figures~\ref{fig:mx-pm-single} and~\ref{fig:mx-pm-te} show the corresponding
afternoon and evening net flows for the intervals 15:00--18:00 and
15:00--21:00.
Relative to the morning period, the dominant flow directions partially reverse:
many central locations act as net sources rather than sinks, and outward-directed
flows toward peripheral areas become more prominent.
This outward redistribution becomes more pronounced as the time-elapsed window
is extended, with additional peripheral destinations entering the top $75^{\text{th}}$ percentile.

However, the afternoon and evening patterns are notably less spatially coherent
than the morning inflow.
Flows are more dispersed in direction, and the radial structure is weaker,
reflecting a mixture of return commuting, localized circulation, and residual
central activity.

\noindent\emph{Interpretation.}

In contrast to the synthetic case, the aggregated data do not exhibit
a stable ordering or recurrence structure across days.
The synthetic dynamics show limited stochastic variability due to
trajectory sampling; both the set of high effective-distance pairs and their
effective-distance magnitudes remain far more stable than in the aggregated
setting. We note that this difference is partly a consequence of the deliberately imposed
daily periodicity in the synthetic generator.
The same framework could be modified to enforce longer periodic cycles (e.g.,
weekly), which would be expected to produce more heterogeneous recurrence
structure more comparable to aggregated data.

Overall, these results illustrate both the utility and the limitations of
time-elapsed net flows when applied to strongly aggregated OD data.
While coarse temporal resolution obscures fine-grained movement structure,
time-elapsed net flows nonetheless provide a compact, directional summary of
collective redistribution over extended periods, enabling qualitative
comparison across time windows and between synthetic and empirical settings.

\section{Time-elapsed OD distance traveled and effective distance}
\label{sec:tedist}

We now extend the time-elapsed framework beyond trip counts and consider
\emph{how far} a PEP travels between an
OD pair over a time-elapsed window.
Unlike single-interval distances, which record only direct transitions,
time-elapsed distances incorporate all admissible intermediate paths
connecting an origin to a destination, weighted by their probabilities under
the underlying mobility dynamics.

This construction captures situations in which travel between two locations
systematically deviates from the most direct route, for example due to
detours, transfers, congestion, or structural constraints imposed by the
mobility network.
To compare such effects across OD pairs with widely different geographic
separations, we introduce an OD-specific normalization, which we refer to as
the \emph{effective distance}.

Throughout this section we focus on travel distance; an identical construction
applies to travel time.

\subsection{Time-elapsed OD distance}
\label{subsec:tedistcalc}

Fix an OD pair $(j,i)$.
We consider all admissible paths that originate at $j$ at the beginning of a
time-elapsed window and first reach $i$ at a later time-step, without passing
through either $j$ or $i$ at intermediate times (see Appendix~\ref{appdx:tedistcalc}
for the formal recursion).

Let $x^t_{ij}$ denote the mean distance traveled by paths that first reach
destination $i$ exactly at time-step $t$, and let $\pi^t_{ij}$ denote the
probability of such paths.
Both quantities are computed recursively from the single-step transition
matrices together with the observed single-step trip distances; the full
recursive construction is provided in Appendix~\ref{appdx:tedistcalc}.

The mean distance traveled from $j$ to $i$ over a window
$t_1 \le t \le t_2$ is defined as the probability-weighted average
\begin{equation}
\bar{x}^{t_1,t_2}_{ij}
=
\frac{\sum_{t=t_1}^{t_2} x^t_{ij}\,\pi^t_{ij}}
     {\sum_{t=t_1}^{t_2} \pi^t_{ij}} .
\label{eq:tedist}
\end{equation}
where  the denominator is the total probability mass associated with all
admissible return paths over the time-elapsed window.
For an OD pair $(i,j)$, this quantity is given by
\[
P^{t_1,t_2}_{ij} = \sum_{t=t_1}^{t_2} \pi^t_{ij},
\]
where $\pi^t_{ij}$ is the probability that a path first reaches $i$ from $j$ at
time-step $t$.

The above expression for $\bar{x}^{t_1,t_2}_{ij}$ represents the expected path length between $j$ and $i$
induced by the mobility dynamics over the time-elapsed window.
When the window spans a single time-step, it reduces to the usual mean
single-interval trip distance.

We will refer to $P^{t_1,t_2}_{ij}$ as the \emph{total path-hit probability} of the OD pair
over the window, and use it as a diagnostic measure of how strongly a given effective-distance
connection is supported by the underlying dynamics.

\subsection{Normalization and effective distance}
\label{subsec:teeffdistcalc}

Raw time-elapsed distances are not directly comparable across OD pairs, since
geographic separation varies substantially. We therefore normalize each time-elapsed distance by an OD-specific baseline,
defining the \emph{effective distance}. The term \emph{effective distance} is used here in a descriptive,
OD-specific sense and should not be confused with the effective distance
introduced in epidemic spreading on networks~\cite{bh:hdcn}.
While both constructions normalize path length by network structure,
the present definition is based on time-elapsed mobility paths inferred
from aggregated OD flows rather than on arrival-time geometry of contagion.

A natural baseline is the typical trip distance for the OD pair.
However, not all OD pairs realized through time-elapsed paths appear in the
single-step data.
We therefore impute baseline distances using a regression between geographic
distance and observed median trip distance.

Specifically, we perform a weighted linear regression (weighted by trip counts; dominating weight is for self-loops)
between geographic distance and median observed trip distance for all OD pairs
present in the data.
The fitted relation is used to estimate a baseline distance for OD pairs absent
from the data.
To remain conservative, normalization is performed using the baseline distance
plus one standard deviation (or the regression \gls{RMSE} when imputed).

The effective distance for OD pair $(i,j)$ over window $(t_1,t_2)$ is defined as
\[
D^{\mathrm{eff}}_{ij}
=
\frac{\bar{x}^{t_1,t_2}_{ij}}
     {\text{baseline}_{ij} + \sigma_{ij}} .
\]
Values of effective distance substantially larger than one indicate travel that is systematically
longer than expected given geographic separation, signaling indirect,
constrained, or inefficient movement between the OD pair.

\subsection{Aggregated data baseline construction}
\label{subsec:teeffdist-baseline}

We apply the effective-distance construction to aggregated OD data from the
NetMob 2024 Data Challenge, focusing on the Mexico City metropolitan region and
weekday morning commuting windows (06:00--09:00--12:00) throughout 2019.

\begin{figure}[H]
\centering
\includegraphics[width=0.65\linewidth]{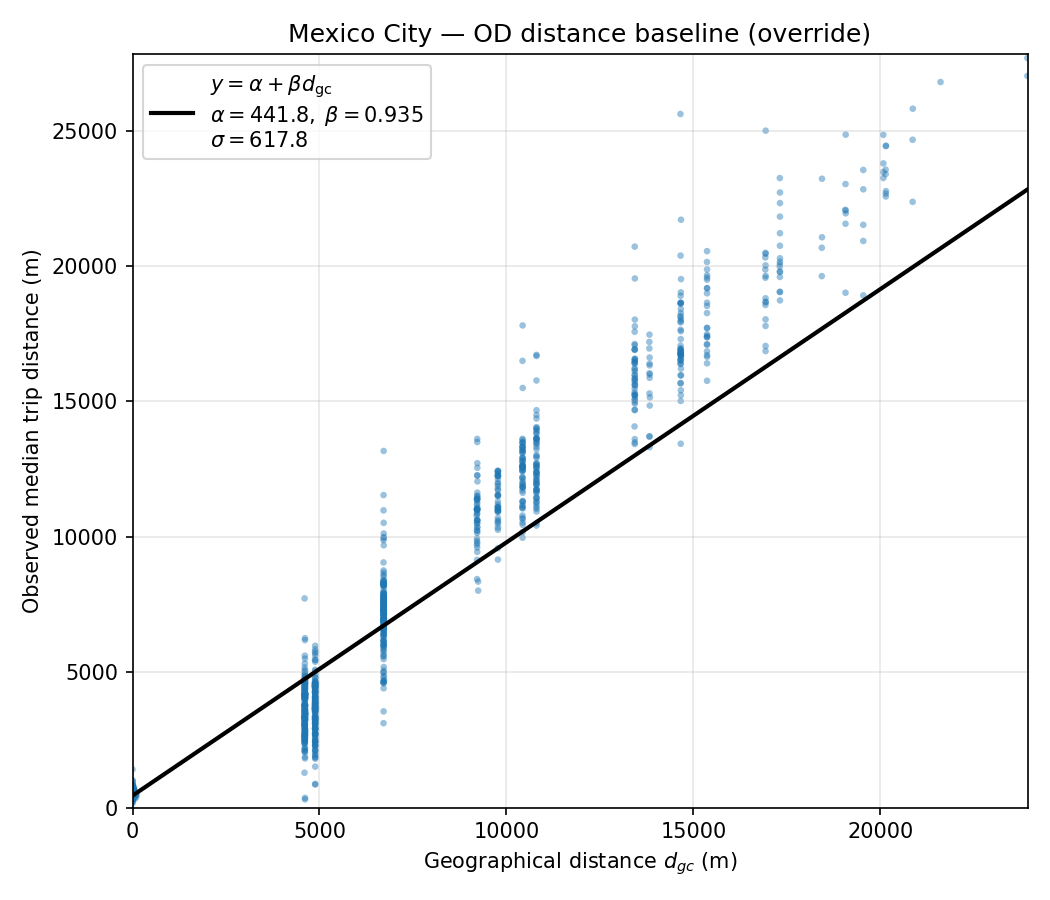}
\caption{
Mexico City: weighted linear fit of observed median trip distance versus
geographic distance, used to construct OD-specific effective-distance
baselines.
}
\label{fig3}
\end{figure}

Figure~\ref{fig3} shows the weighted regression between geographic distance and
observed median trip distance.
While the fit exhibits substantial variance, it captures the overall trend and
provides a workable normalization baseline for effective distance.

All spatial results and path decompositions are discussed in
Section~\ref{subsec:teeffdistres}.
Supplementary diagnostics, including temporal persistence of high effective
distance OD pairs, are provided in Appendix~\ref{app:tedist-timesweep}.

\subsection{Results: effective distance and path structure}
\label{subsec:teeffdistres}

We now examine time-elapsed effective distances in both synthetic and aggregated
settings, focusing on \emph{globally-unconnected pairs} (GUPs), defined as OD
pairs that are not connected by a direct single-step transition.
Such pairs provide a stringent test of whether effective distance captures
structurally indirect travel, rather than merely reflecting geographic separation
or the presence of a known direct route.

\noindent\emph{Synthetic data:}

We illustrate the interpretation of effective distance using a synthetic example
from the Atlanta network.
We consider the time window 06:00--08:00 on 2025--06--01, corresponding to four
30-minute time steps.

\begin{figure}[H]
\centering
\includegraphics[width=0.6\linewidth]{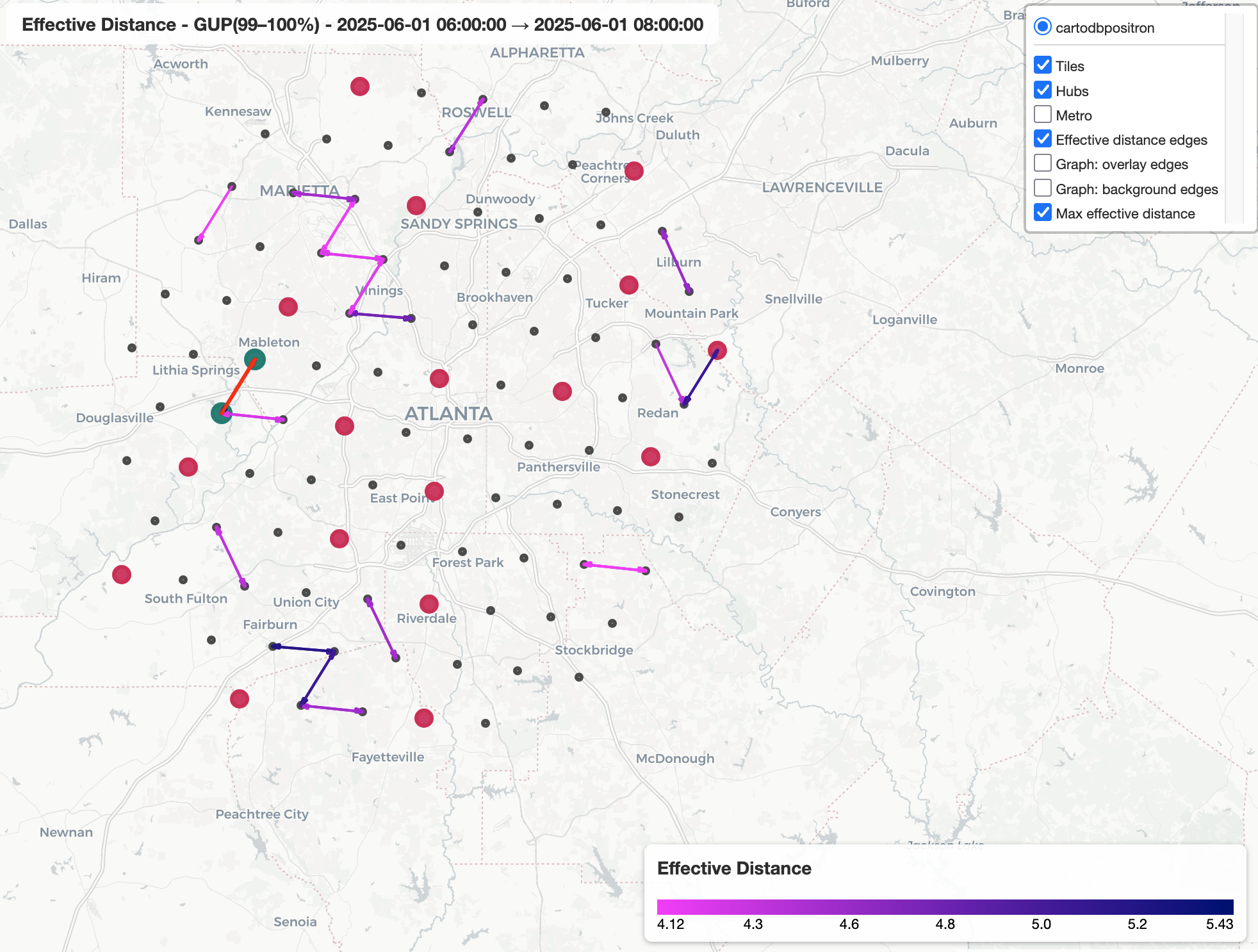}
\caption{
Synthetic effective distances for GUPs,
2025--06--01 06:00--08:00.
The largest effective distance is highlighted in green.
}
\label{fig:synth-eff-map}
\end{figure}

Figure~\ref{fig:synth-eff-map} shows effective distances for
GUPs in the top $99^{\text{th}}$ percentile for that window.
The largest value observed is $D^{\mathrm{eff}}_{\text{max}}= 5.36$.
Although the corresponding origin and destination are geographically close,
their effective distance is large, indicating a strong topological separation
in the network.

\begin{figure}[H]
\centering
\includegraphics[width=0.6\linewidth]{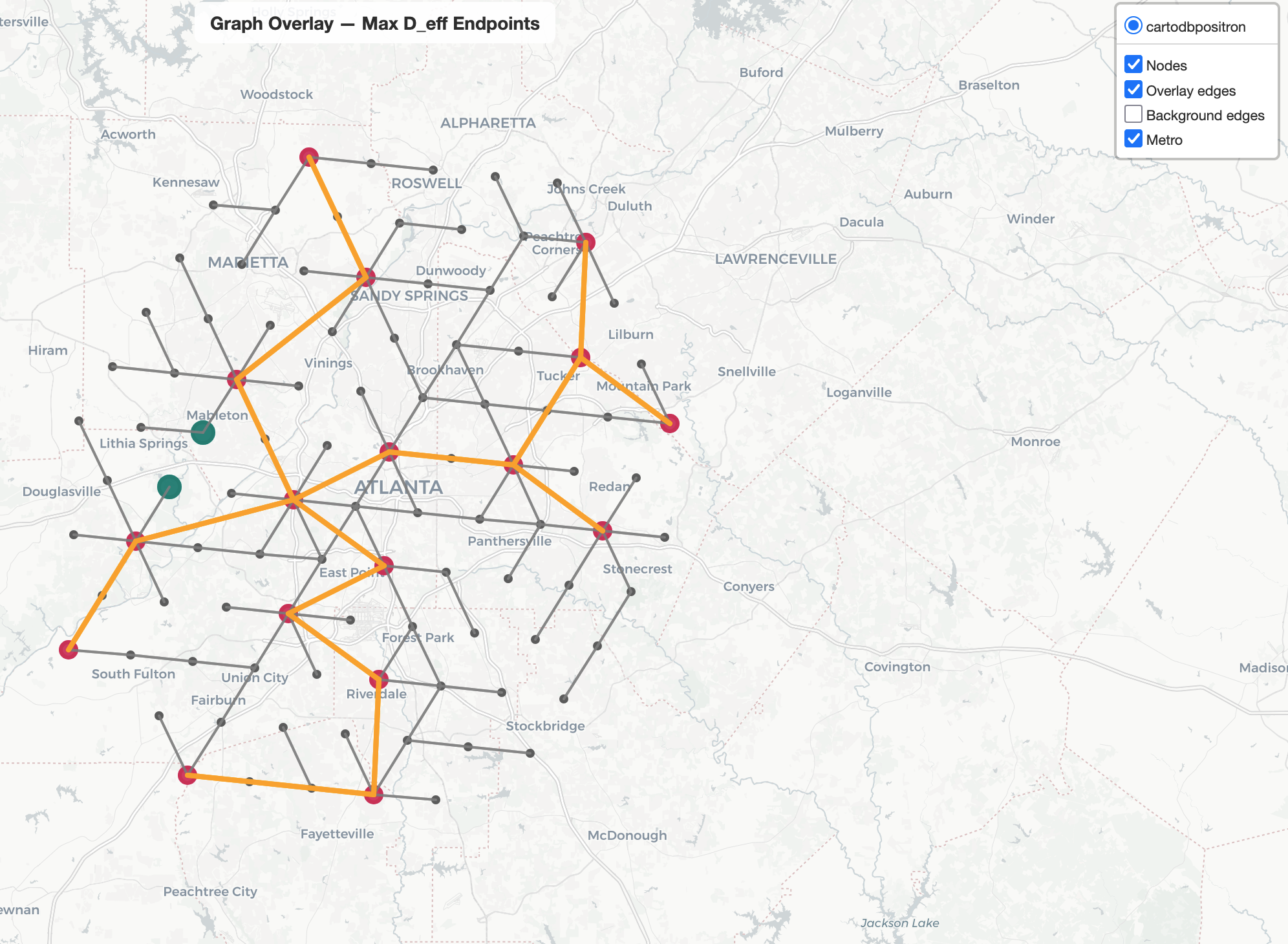}
\caption{
Overlay network highlighting the endpoints of the maximum effective-distance
OD pair.
The absence of a direct connection confirms that the pair is globally
unconnected.
}
\label{fig:synth-eff-overlay}
\end{figure}

To confirm the structural origin of this separation,
Figure~\ref{fig:synth-eff-overlay} shows the overlay network with the endpoints
of the maximum effective-distance pair emphasized.
Despite their spatial proximity, no direct edge connects these two locations,
and they therefore constitute a GUP.

\begin{figure}[H]
\centering
\includegraphics[width=0.4\linewidth]{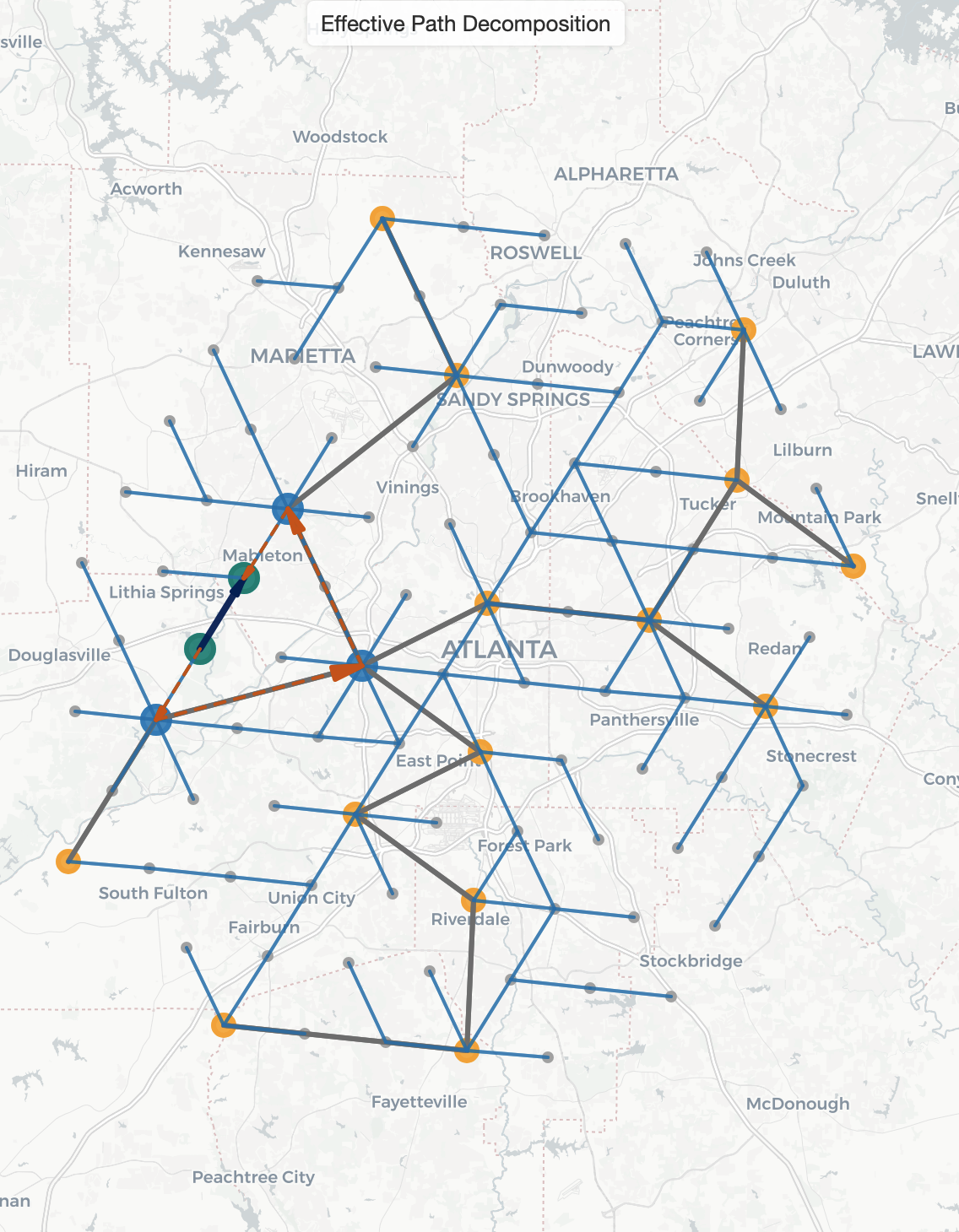}
\caption{
Effective-path decomposition for the maximum effective-distance OD pair.
The dominant contribution follows a four-hop route via intermediate hubs,
including two high-capacity metro-edge segments.
}
\label{fig:synth-eff-path}
\end{figure}

The effective-path decomposition for this OD pair is shown in
Figure~\ref{fig:synth-eff-path}.
Rather than a direct transition, flow is routed through a sequence of
intermediate nodes, resulting in a four-hop path.
Two of these segments correspond to high-capacity metro edges, which dominate
the probability-weighted contribution to the effective distance.

This example illustrates the role of effective distance as a detector of
structural gaps in the network.
Despite short geographic separation, the absence of a direct connection forces
travel to proceed through multiple intermediates.
The resulting large effective distance therefore reflects the topology and
capacity structure of the network and its time-dependent dynamics, rather than
spatial separation alone or aggregation artifacts.

To quantify how strongly this OD connection is supported by the underlying
dynamics, we also report the total probability mass associated with all
admissible return paths over the time-elapsed window.
For an OD pair $(i,j)$, this quantity is given by
\[
P^{t_1,t_2}_{ij} = \sum_{t=t_1}^{t_2} \pi^t_{ij},
\]
where $\pi^t_{ij}$ is the probability that a path first reaches $i$ from $j$ at
time-step $t$.

For the synthetic OD pair attaining the maximum effective distance,
the total path probability over the window is
\[
P^{06{:}00,08{:}00}_{\text{max}} = 2.09 \times 10^{-4}.
\]
Thus, while the effective distance is large, the corresponding connection is
realized only by a small fraction of the population, consistent with the
interpretation that large effective distance reflects structurally indirect but
low-probability routing through the network.

\noindent\emph{Aggregated data:}

\begin{figure}[H]
\centering
\includegraphics[width=0.6\linewidth]{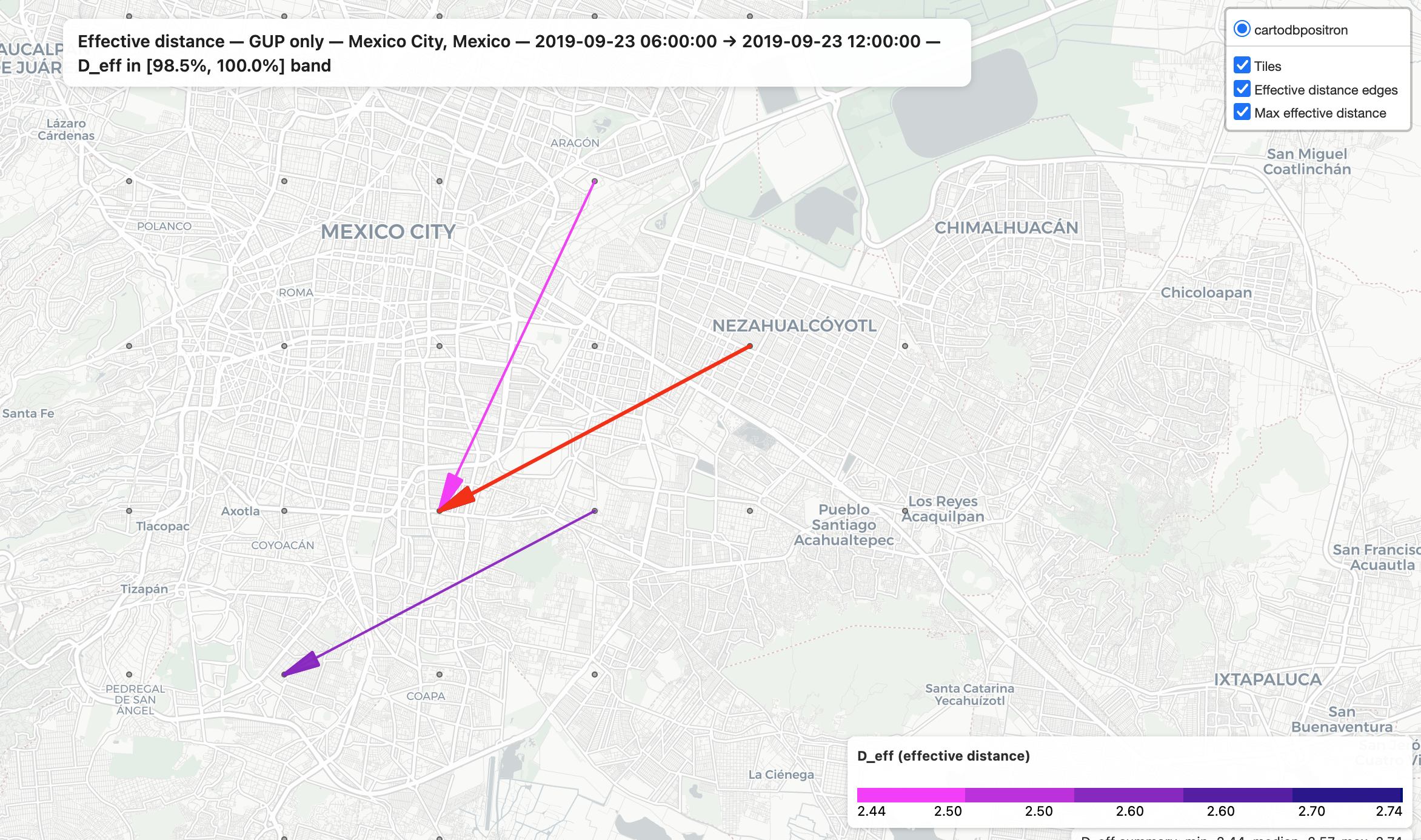}
\caption{
Mexico City:
effective distances for GUPs,
2019--09--23 06:00--12:00.
The maximum effective distance is $D_{\mathrm{eff}}=2.67$.
}
\label{fig:agg-eff-map}
\end{figure}

We now contrast this behavior with aggregated OD data from Mexico City.
Figure~\ref{fig:agg-eff-map} shows effective distances for  \gls{GUP}  OD pairs over the
time window 06:00--12:00 on 2019--09--23, corresponding to two three-hour
intervals. Here and throughout, alphanumeric labels such as \texttt{9g3w6} denote geohash
cell identifiers used in the NetMob dataset, and strings such as \texttt{8644c1a77ffffff} denote
H3 hierarchical hexagonal indices at resolution~6.

The largest effective distance observed is $D^{\mathrm{eff}}_{\text{max}}= 2.67$, associated
with the OD pair \texttt{9g3w6} $\rightarrow$ \texttt{9g3w0}.

\begin{figure}[H]
\centering
\includegraphics[width=0.6\linewidth]{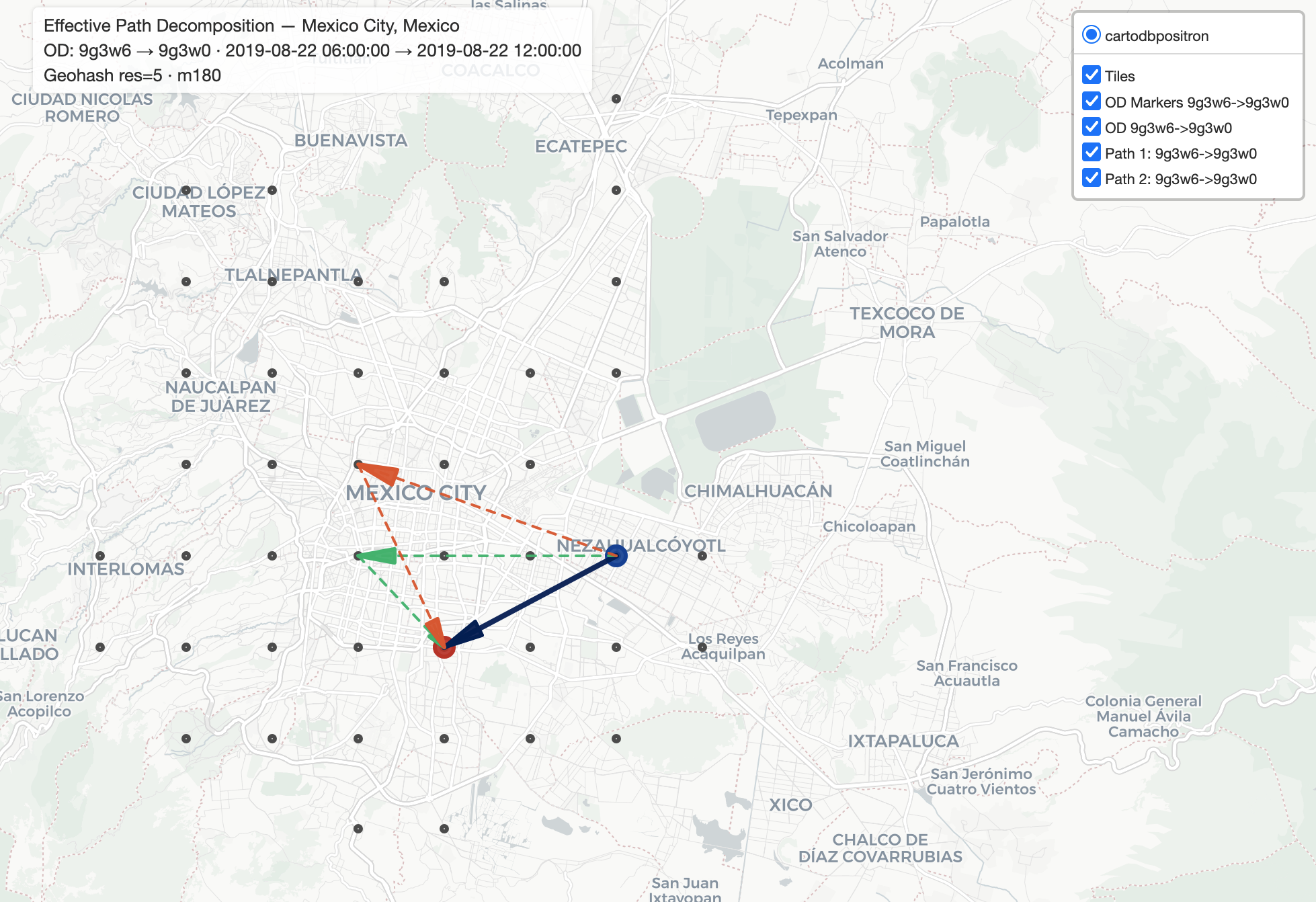}
\caption{
Aggregated-data effective-path decomposition for the OD pair
\texttt{9g3w6} $\rightarrow$ \texttt{9g3w0}.
Two distinct paths contribute to the effective distance.
}
\label{fig:agg-eff-path}
\end{figure}

In this setting, the underlying network structure is unknown and paths can only
be inferred indirectly from aggregated flows.
The effective-path decomposition in Fig.~\ref{fig:agg-eff-path} shows that this
OD pair is realized through two distinct two-step routes,
\begin{equation*}
\texttt{9g3w6} \rightarrow \texttt{9g3qx} \rightarrow \texttt{9g3w0},
\qquad
\texttt{9g3w6} \rightarrow \texttt{9g3qr} \rightarrow \texttt{9g3w0}.
\end{equation*}
The time-elapsed distance is computed as a probability-weighted average over
these paths, after which the effective distance is obtained by normalization
using an imputed baseline derived from geographic-distance regression.

As in the synthetic case, we compute the total probability mass associated with
all admissible paths supporting this OD connection.
Over the window 06:00--12:00, the aggregated-data OD pair with maximum effective
distance has total path probability
\[
P^{06{:}00,12{:}00}_{\text{max}}  = 1.17 \times 10^{-3}.
\]

Compared to the synthetic example, this probability mass is larger, which is
expected because the dominant contributing routes here are only two steps long,
whereas the synthetic $D^{\mathrm{eff}}_{\text{max}}$ pair is realized primarily via
a four-hop route.
Shorter admissible connections typically accumulate more probability mass under
products of stochastic transition operators.

In addition, the coarser three-hour aggregation can further increase the
effective probability of reaching the destination within the window, because
each observed transition summarizes heterogeneous within-interval movements.
Accordingly, elevated effective distance in the aggregated setting may reflect
a superposition of multiple plausible intermediate realizations consistent with
the same coarse OD flows, rather than a single sharply constrained route.

\noindent\emph{Comparison and role of temporal resolution:}

The contrast between synthetic and aggregated results highlights the critical
role of temporal resolution in effective-distance inference.
In the synthetic setting, 30-minute resolution allows individual multi-hop
routes to be resolved explicitly, so large effective distances can be
unambiguously attributed to specific network-imposed detours. This ambiguity is a known limitation of OD-based inference under strong
temporal aggregation, where multiple plausible paths may be consistent with
the same observed flow~\cite{bbgj:hmma}.

In the aggregated data, transitions are observed at three-hour resolution.
Within such coarse intervals, multiple distinct movement patterns may be
collapsed into a single OD observation, leading to path degeneracy.
As a result, elevated effective distances in the aggregated setting often arise
from the superposition of several plausible intermediate paths rather than a
single dominant route.

This ambiguity is therefore not a failure of the effective-distance
construction.
Rather, it reflects limitations imposed by coarse temporal aggregation.
At higher temporal resolution, effective distance would be expected to isolate
inefficient or indirect routes more sharply, in closer analogy with the
synthetic case.

Overall, effective distance serves its intended purpose as a probe of
\emph{structural inefficiency} in mobility networks: it highlights OD pairs for
which observed movement systematically deviates from direct or expected paths,
revealing detours, bottlenecks, and constraints that are not visible from trip
counts alone.

\section{Time-elapsed return-to-origin distance}
\label{sec:terto}

As a complementary summary statistic to time-elapsed OD
flows and effective distance, we consider the
\emph{return-to-origin} (RTO) distance.
RTO measures the distance traveled by a PEP
before returning to their origin within a specified time-elapsed window.
When averaged across origins, RTO provides an intrinsic, population-level
characterization of mobility scale, analogous in spirit to radius-of-gyration
measures but derived from time-elapsed flow dynamics.

RTO is a special case of the time-elapsed OD distance construction introduced in
Section~\ref{sec:tedist}, obtained by setting the destination equal to the
origin.
Throughout this section we focus on distance-based RTO; time- and speed-based
variants can be obtained by the same construction with time-valued edge costs.

\subsection{Definition and variants}

RTO distance is a special case of the time-elapsed
OD distance in which the destination coincides with the
origin.
Let $\bar{x}^{t_1,t_2}_{ij}$ denote the mean time-elapsed distance from origin
$j$ to destination $i$, as defined in Eq.~\eqref{eq:tedist} (see Appendix~\ref{appdx:tedistcalc}).
Setting $i=j$ yields the mean RTO distance for location $j$ over the window
$t_1 \le t \le t_2$,
\begin{equation}
\bar{x}^{t_1,t_2}_{jj}
=
\frac{\sum_{t=t_1}^{t_2} x^t_{jj}\,\pi^t_{jj}}
     {\sum_{t=t_1}^{t_2} \pi^t_{jj}},
\label{eq:rto-def}
\end{equation}
where $x^t_{jj}$ and $\pi^t_{jj}$ are the mean distance and probability of paths
that first return to $j$ at time-step $t$.
The recursive computation of these quantities is summarized in
Appendix~\ref{appdx:tedistcalc}.

We distinguish two RTO variants that separate near-origin stationarity from
true excursions:

\begin{itemize}
\item \emph{Home RTO.}
Restricts attention to paths that return to the origin at the first time-step.
In this case,
\begin{equation}
\bar{x}^h_j = \bar{x}^{1,2}_{jj} = d^1_{jj},
\end{equation}
where $d^1_{jj}$ is the (median) distance traveled during the self-transition
at location $j$.
Home RTO therefore captures short-range, intra-cell mobility at the spatial
resolution of the data.

\item \emph{Roaming RTO.}
Excludes the first self-transition and conditions on paths that leave the origin
before returning.
The roaming RTO for location $j$ is
\begin{equation}
\bar{x}^r_j = \bar{x}^{2,t_2}_{jj},
\end{equation}
capturing commuting- or excursion-scale mobility.
\end{itemize}

City-level RTO values are obtained by averaging over origin locations using
weights induced by the aggregated OD flows at the initial time-step.

The probability that a generic PEP is located at origin $j$ at the beginning of
the window is
\begin{equation}
p^1_j
=
\frac{\sum_i f^1_{ij}}{\sum_{i,j} f^1_{ij}},
\end{equation}
where $f^1_{ij}$ denotes the observed flow from $j$ to $i$ at the first time-step. As in the net-flow construction, these weights are estimated directly from
observed aggregated flows at the initial time-step rather than from a fixed-point
distribution.
This choice ensures that RTO measures remain data-driven and applicable even
when a stationary distribution is not well defined.

The relative distribution of PEPs that leave their origin at the first step is
\begin{equation}
p^r_j
=
\frac{\sum_{i \neq j} f^1_{ij}}{\sum_{k}\sum_{i \neq k} f^1_{ik}},
\end{equation}
which defines the weights used to compute the city-level \emph{roaming RTO}:
\begin{equation}
\bar{x}^r
=
\sum_j p^r_j \, \bar{x}^{2,t_2}_{jj}.
\end{equation}

Similarly, the relative distribution of PEPs that remain at their origin at the
first time-step is
\begin{equation}
p^h_j
=
\frac{f^1_{jj}}{\sum_k f^1_{kk}},
\end{equation}
and the city-level \emph{home RTO} is
\begin{equation}
\bar{x}^h
=
\sum_j p^h_j \, d^1_{jj}.
\end{equation}

\subsection{Synthetic data}

We first examine RTO behavior in the synthetic network.
As expected, home RTO is substantially smaller than roaming RTO, reflecting
the distinction between near-origin stationarity and excursion-scale mobility.
However, home RTO is not vanishingly small: because distances are measured at
the spatial resolution of the underlying H3 grid, self-transitions correspond
to movement within a cell rather than zero-length trips.

As a result, synthetic home RTO reflects typical intra-cell displacement at the
chosen spatial resolution, yielding distances on the order of a single cell
diameter.
This behavior is consistent with the interpretation of home RTO as capturing
short-range, near-origin mobility rather than literal immobility.

\begin{figure}[H]
\centering
\includegraphics[width=0.85\linewidth]{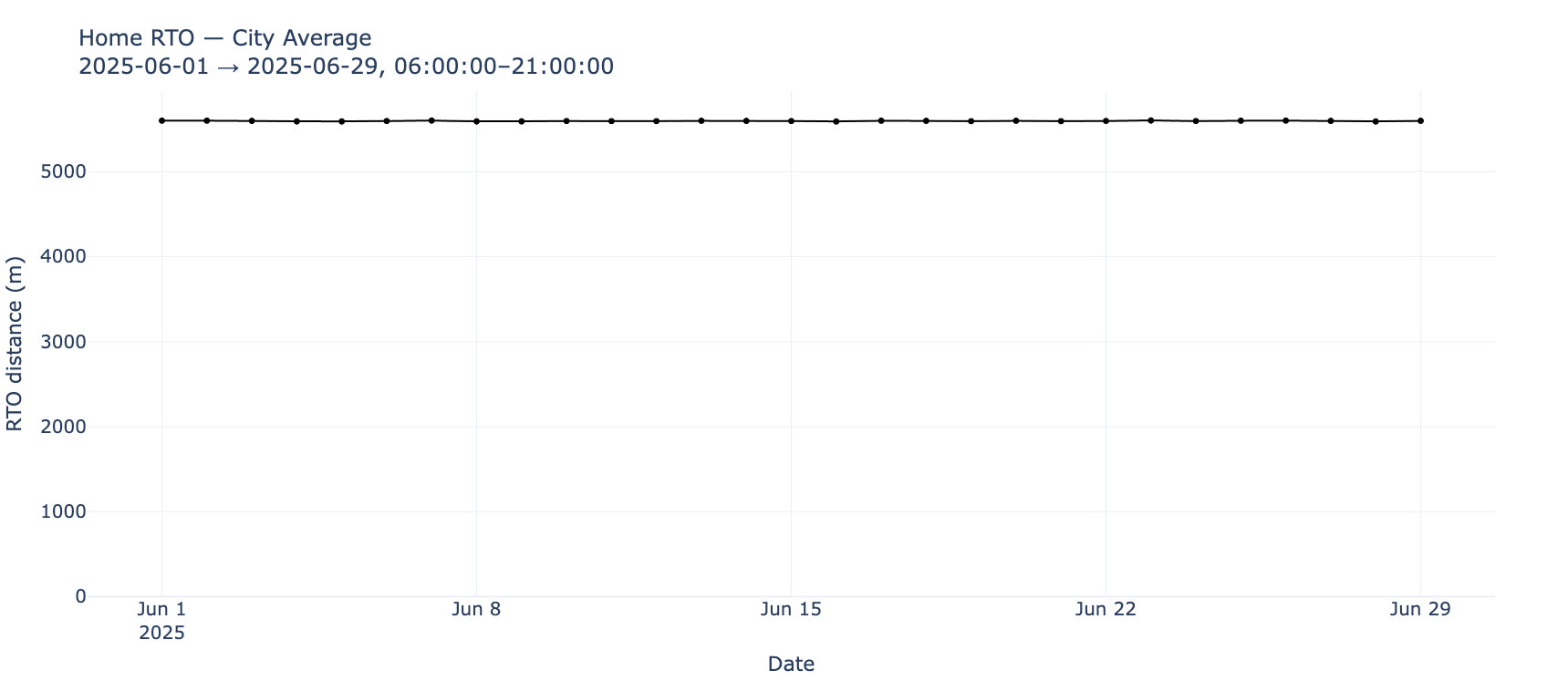}
\caption{
Synthetic home RTO, city average,
2025--06--01 to 2025--06--29.
The vertical axis starts at zero to emphasize the absolute scale.
}
\label{fig:rto-synth-home}
\end{figure}

Figure~\ref{fig:rto-synth-home} shows the synthetic \emph{home RTO}
(city average) over 2025--06--01 to 2025--06--29.
The plot is shown with a zero baseline to emphasize the absolute scale of home
RTO relative to roaming RTO.
On this scale, day-to-day variation appears visually small.

\begin{figure}[H]
\centering
\includegraphics[width=0.85\linewidth]{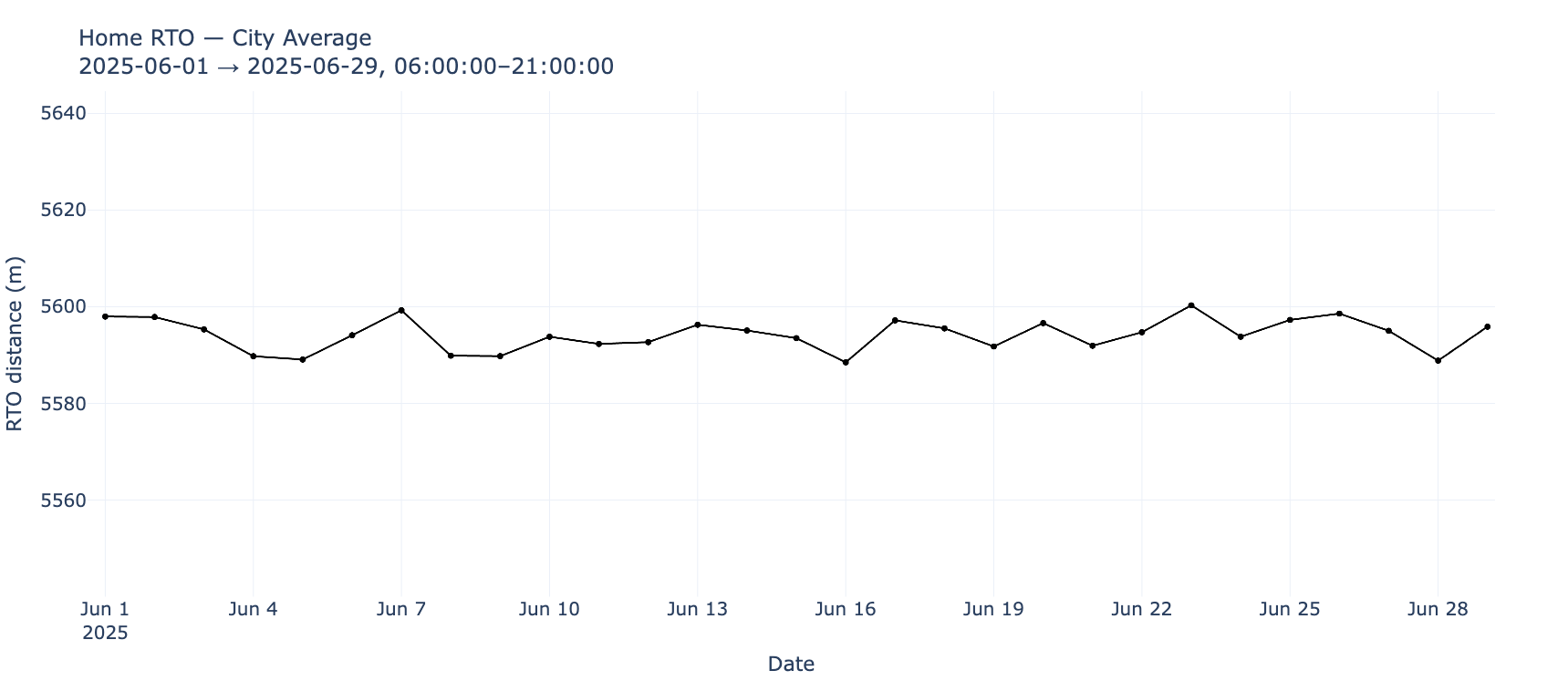}
\caption{
Synthetic home RTO, city average, shown on a restricted vertical scale.
Small but persistent day-to-day fluctuations reflect intrinsic stochastic
exploration of the network and finite-sample effects in trajectory realization.
}
\label{fig:rto-synth-home-var}
\end{figure}

To make these variations explicit, Figure~\ref{fig:rto-synth-home-var} shows the
same time series with a restricted vertical axis.
On this scale, small but systematic fluctuations are clearly visible.
In this controlled setting, variability arises from two sources:
(i) intrinsic stochasticity of the time-dependent Markov dynamics, which permit
randomized direction and path exploration even under a fixed transition
schedule, and
(ii) finite-sample effects associated with stochastic trajectory realizations
in the PEP simulation.

Because transitions allow probabilistic exploration of neighboring locations,
short-range return distances fluctuate over time even in the absence of
structural or calendar-driven changes.
These fluctuations reflect diffusion-like behavior inherent to the model rather
than aggregation or measurement effects.

\begin{figure}[H]
\centering
\includegraphics[width=0.85\linewidth]{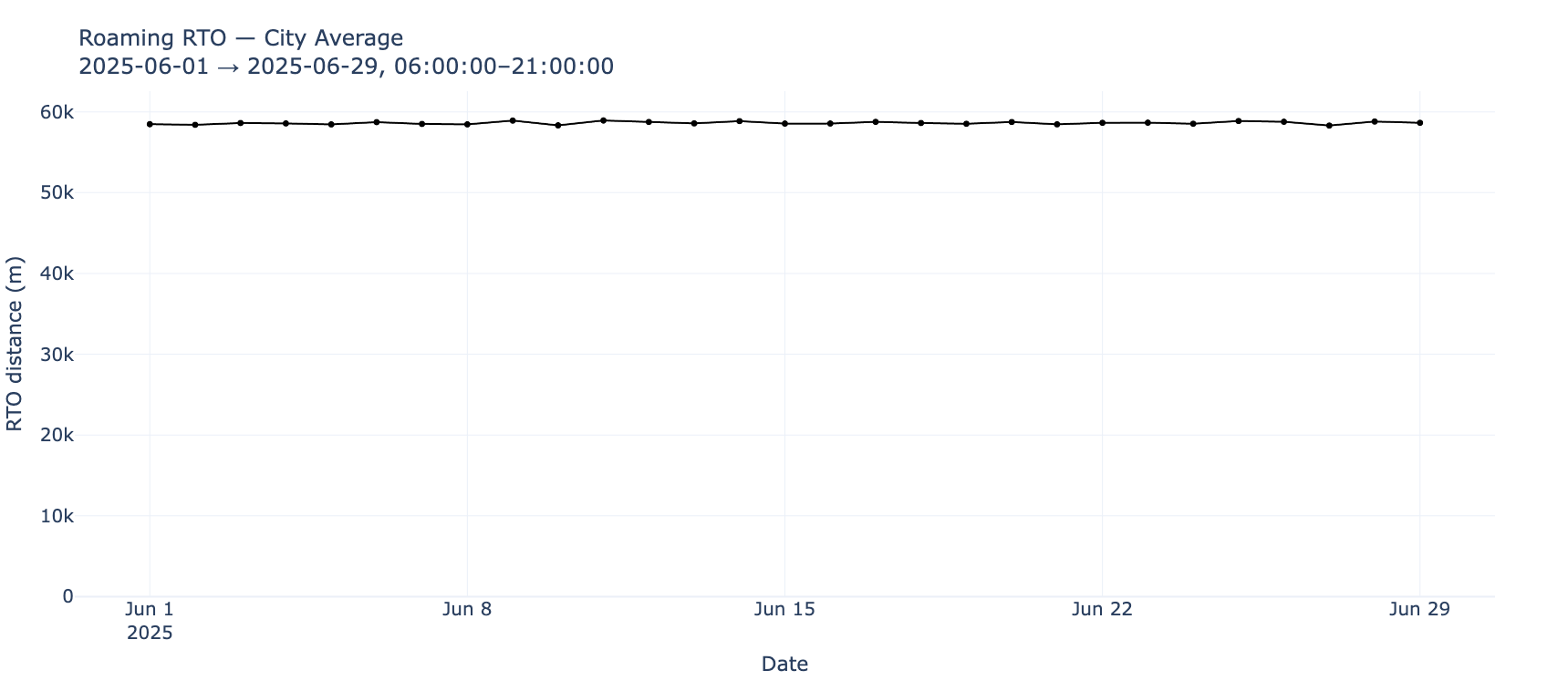}
\caption{
Synthetic roaming RTO, city average,
2025--06--01 to 2025--06--29.
}
\label{fig:rto-synth-city}
\end{figure}

Figure~\ref{fig:rto-synth-city} shows the corresponding synthetic
\emph{roaming RTO} for June~2025, again plotted with a zero baseline to emphasize
its absolute scale relative to home RTO.
Roaming RTO is much larger than home RTO, consistent with longer excursions
before return.
On this scale, day-to-day variability appears modest.

\begin{figure}[H]
\centering
\includegraphics[width=0.85\linewidth]{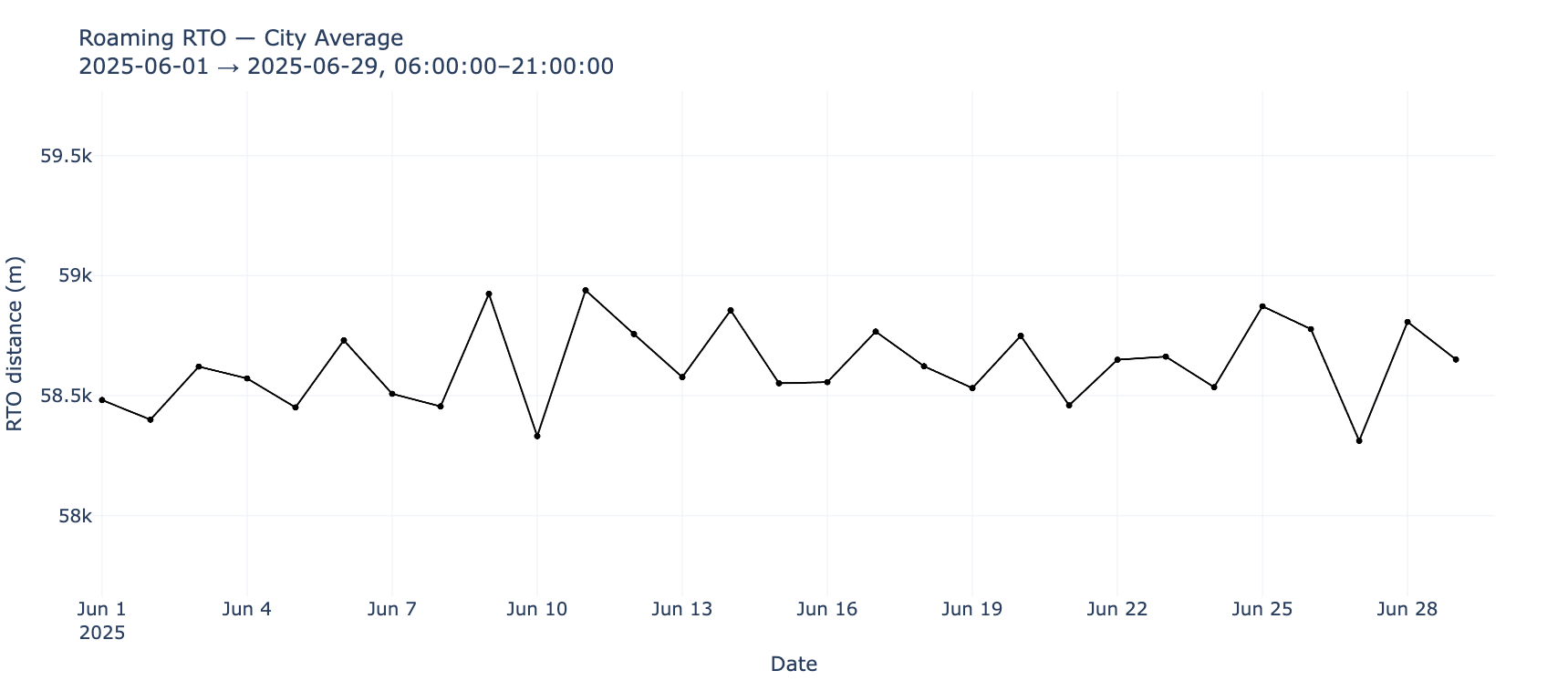}
\caption{
Synthetic roaming RTO, city average, shown on a restricted vertical scale.
Relative fluctuations are larger than for home RTO, reflecting stochastic
path exploration over longer excursion distances under fixed Markov dynamics.
}
\label{fig:rto-synth-city-var}
\end{figure}

Figure~\ref{fig:rto-synth-city-var} shows the same roaming RTO time series with a
restricted vertical axis.
Compared to home RTO, relative fluctuations are more pronounced, reflecting the
greater sensitivity of long excursion paths to stochastic routing choices,
directional dispersion, and finite-sample effects in the PEP simulation.
As in the home case, these variations arise under a fixed transition schedule
and do not reflect structural or calendar-driven changes.


This mechanism mirrors the limited stochastic variability observed in synthetic
effective-distance time sweeps, reinforcing the interpretation that both
measures reflect stable underlying dynamics perturbed by probabilistic path
exploration.

\subsection{Aggregated data: Mexico City}

We now compute RTO from aggregated OD data for Mexico City over 2019 using the
time-elapsed window 06:00--21:00.
As in the synthetic case, home RTO is much smaller than roaming RTO, reflecting
the distinction between near-origin stationarity and excursion-scale mobility.

\begin{figure}[H]
\centering
\includegraphics[width=0.85\linewidth]{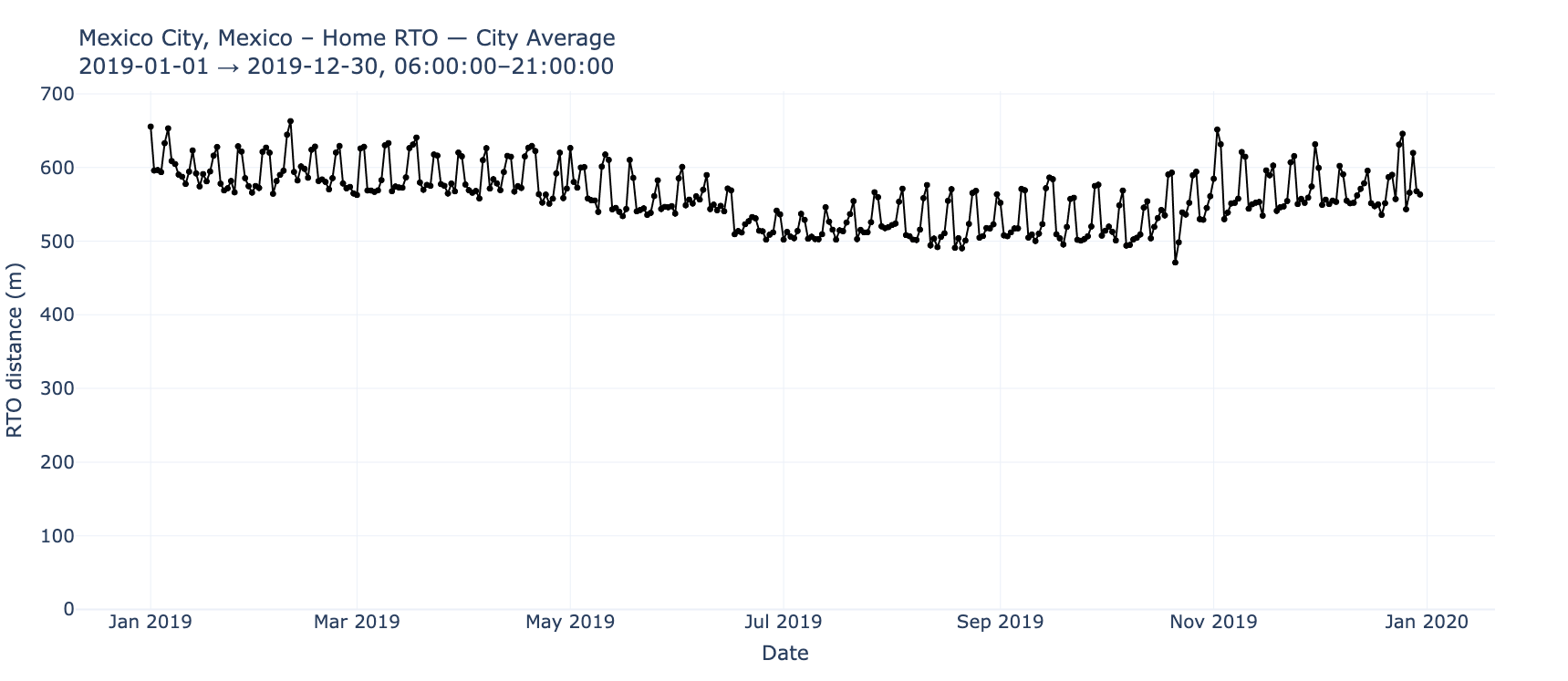}
\caption{
Mexico City home RTO, city average, 2019.
}
\label{fig:rto-agg-home}
\end{figure}

Figure~\ref{fig:rto-agg-home} shows the \emph{home RTO} (city average).
Home RTO varies over time, but the weekday--weekend modulation is comparatively
muted, consistent with the interpretation that near-origin movement is less
sensitive to commuting schedules than longer excursions. This contrast mirrors the effective-distance diagnostics: measures dominated by
local or near-origin movement are less sensitive to calendar-driven mobility
patterns than those capturing longer excursions.

\begin{figure}[H]
\centering
\includegraphics[width=0.85\linewidth]{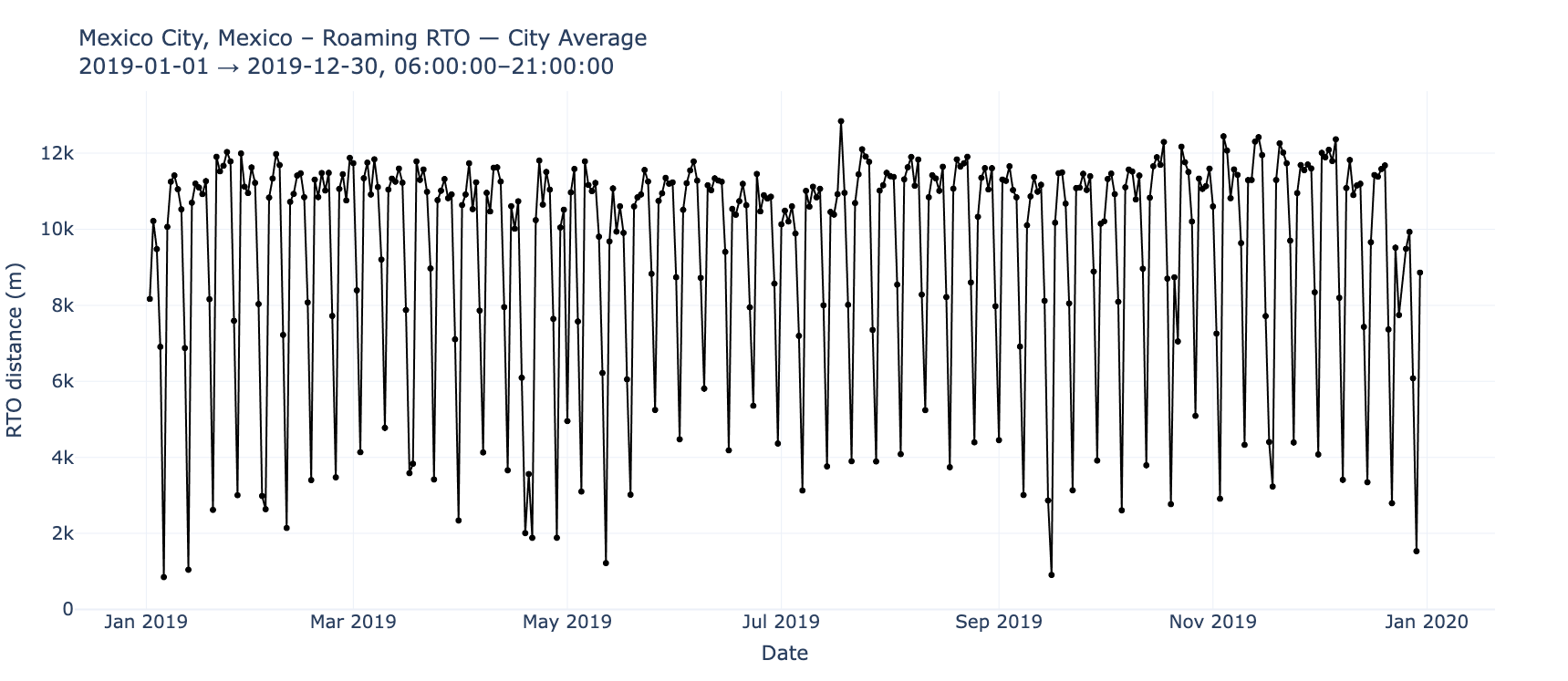}
\caption{
Mexico City roaming RTO, city average, 2019.
}
\label{fig:rto-agg-city}
\end{figure}

Figure~\ref{fig:rto-agg-city} shows the corresponding \emph{roaming RTO}.
Here the weekday--weekend signature is much more pronounced, consistent with
roaming RTO capturing commute-like excursions and longer out-and-back trips.
In addition, both home and roaming RTO exhibit slower variations across the
year.
These may reflect seasonal effects, changing activity patterns, or other
contextual factors (e.g., holidays and events).
We do not attempt to attribute these variations causally in this work.

Finally, the aggregated setting introduces measurement effects absent in the
synthetic model.
With three-hour time steps, multiple within-interval movements are compressed.
Distance statistics must therefore be derived from reported trip-distance summaries (e.g., medians) rather than controlled edge-length means.
These factors can amplify apparent variability and blur fine-scale temporal
structure, particularly for roaming RTO, which is driven by longer excursions.

A detailed empirical analysis of RTO behavior across cities, temporal regimes,
and aggregation scales is left for future work.
The definitions provided here demonstrate that such measures are naturally
supported by the time-elapsed framework and can be computed without access to
individual trajectories.

\section{Conclusion}
\label{sec:conc}

This paper addresses the problem of extracting longer-term mobility structure
from space- and time-aggregated collective OD data.
Starting from a pseudo Markov-chain representation of aggregated flows, we develop
a consistent framework for constructing time-elapsed mobility measures,
including net OD flows, time-elapsed distance, effective distance, and
RTO quantities, without relying on individual trajectories
or path-level observations.

Despite the well-known mobility-aliasing effects induced by coarse temporal
aggregation in datasets such as NetMob 2024, we show that meaningful structure
can still be recovered.
Time-elapsed net OD flows reproduce expected large-scale commuting patterns,
providing a qualitative consistency check for the framework.
These flows are not imposed a priori but emerge from the repeated application
of time-dependent transition operators inferred from aggregated data.

Time-elapsed effective distance offers a complementary perspective by
highlighting OD pairs whose realized travel is systematically
more indirect than expected given their geographic separation.
Through comparisons between synthetic data and aggregated real-world data,
we show that effective distance successfully identifies structurally indirect
connections, including GUPs.
In the synthetic setting, where network structure and temporal resolution are
fully controlled, effective distances are stable and directly attributable to
known routing constraints.
In aggregated data, effective distances reflect a superposition of plausible
paths within coarse time windows, leading to greater variability and ambiguity.
These differences underscore the central role of temporal resolution in
interpreting higher-order mobility measures. This observation is consistent with prior work emphasizing the inherently
multiscale nature of human mobility and the dependence of inferred structure
on the chosen spatial and temporal scales~\cite{aal:shm}.

While effective distance can exhibit apparent spatial and temporal recurrence
in both settings, we emphasize that this work does not perform a formal
persistence or flow analysis.
Observed recurrence should therefore be interpreted as a diagnostic signal
rather than a definitive indicator of structural persistence. Formal persistence and flow-based characterizations require additional
assumptions and dedicated analytical frameworks, which lie outside the scope
of the present work (e.g.,~\cite{cl:dacbdccff}).
Nevertheless, effective distance appears to serve its intended role as a probe
for indirect, inefficient, or constrained mobility, pointing to OD pairs where
detours, transfers, or network limitations may be at play.

RTO distance measures extend the framework to capture the typical
spatial scale of excursions encoded in aggregated dynamics.
Home and roaming RTO distances separate near-origin activity from longer-range
movement and provide a collective analogue of individual radius-of-gyration
concepts.
In aggregated data, RTO measures are influenced by temporal resolution,
summary statistics of trip length, and contextual effects such as weekday--weekend
structure and seasonal variation.
Accordingly, RTO is best viewed here as a first-order descriptive quantity,
with more detailed interpretation deferred to settings with higher-resolution
or complementary data.

At sufficiently large spatial and temporal scales, the existence of stationary
or slowly evolving distributions in the pseudo Markov-chain model suggests the
possibility, in principle,  of identifying long-term drift or relocation of populations, should
data with appropriate resolution and coverage become available. Such collective
signals would not correspond to individual migration decisions, but rather to
system-level reorganization emerging from repeated, structured flows over time.
This perspective resonates, at a qualitative level, with anthropological
accounts in which seasonal mobility, redistribution, and long-horizon movement
play a role in shaping social organization and settlement patterns
(e.g.,~\cite{gw:doe}).

In the context of the NetMob 2024 dataset, a natural application is the
identification of large-scale population drift and structurally constrained
movement patterns.
With additional data sources, such analyses could inform questions relevant to
several Sustainable Development Goals~\cite{un:sdg}, including \gls{SDG}~8 (Decent
Work and Economic Growth), SDG~9 (Industry, Innovation and Infrastructure),
SDG~11 (Sustainable Cities and Communities), and SDG~13 (Climate Action).
Future work may also explore extensions beyond the pseudo-Markov assumption,
incorporating additional constraints or temporal dependencies to better reflect
real-world mobility dynamics. Time-elapsed mobility measures of the type developed here may serve as complementary inputs to applied studies of urban dynamics, accessibility, and infrastructure efficiency, where aggregated data are increasingly used to inform planning and policy~\cite{ysctw:iawsssd,ypm:ilablumddud,wlcqjb:e15mcpaud}.

\section*{Code availability}
All code used to compute the time-elapsed mobility measures reported
in this paper is available as open-source software at
\href{https://github.com/asif-shakeel/Network-Mobility-Measures}
{github.com/asif-shakeel/Network-Mobility-Measures}.

\section*{Acknowledgments}
The authors would like to acknowledge productive participation from the members of David Meyer's research group, in  particular, Itai Maimon for initiating the discussion that led to the RTO definition, Orest Bucicovschi, David Rideout and Jiajie Shi.
 
 Mobility data were provided by Cuebiq and made available through the NetMob 2024 Data Challenge.
All data were anonymized and aggregated prior to release. Cuebiq does not review, endorse,
or take responsibility for the analyses or conclusions presented in this work.

%
%
%
%

\newpage

\appendix

\section{Recursive computation of time-elapsed OD distances}
\label{appdx:tedistcalc}

This appendix provides the technical details underlying the computation of
time-elapsed OD distances used in Section~\ref{subsec:tedistcalc}.
The purpose of this appendix is to make explicit how path probabilities and
path lengths are propagated through time while excluding trivial loops through
the origin and destination.

\subsection{Path probabilities}

Fix an ordered OD pair $(j,i)$.
We track admissible paths that originate at $j$ at time $t=0$ and terminate at
$i$ at some later time-step, without passing through either $i$ or $j$ at any
intermediate time-step. In the following, the superscripts $i,j$ simply keep
track of the OD pair under consideration and can be regarded as bookkeeping.

Let $M^t=[m^t_{kr}]$ denote the time-dependent stochastic transition matrix at
time-step $t$, defined in Eq.~\eqref{eq:Mdef}.
We define the vector
\[
\mathbf{p}^{i,j,t} = [p^{i,j,t}_k],
\]
where $p^{i,j,t}_k$ is the probability that a path starting at $j$ reaches node
$k$ at time-step $t$ without having previously visited either $i$ or $j$.

Initialization is given by
\[
\mathbf{p}^{i,j,1} = [m^1_{kj}],
\]
that is, the $j$-th column of $M^1$.

For $t \ge 2$, probabilities propagate according to
\begin{equation*}
p^{i,j,t}_k
=
\sum_{r \,:\, r \ne i,j} m^t_{kr}\, p^{i,j,t-1}_r .
\end{equation*}

To write this as a matrix operation, define the masked vector
\[
\tilde{\mathbf{p}}^{i,j,t} = [\tilde p^{i,j,t}_k],
\]
where
\[
\tilde p^{i,j,t}_k =
\begin{cases}
p^{i,j,t}_k, & \text{if } k \ne i,j,\\
0, & \text{otherwise}.
\end{cases}
\]
Then
\begin{equation*}
\mathbf{p}^{i,j,t} = M^t\,\tilde{\mathbf{p}}^{i,j,t-1}.
\end{equation*}

Simplifying notation, let
\[
\pi^t_{ij} = p^{i,j,t}_i
\]
denote the probability that a path starting at $j$ first reaches $i$ at time-step $t$, without intermediate returns to $j$.

\subsection{Path distances}

Let $d^t_{kr}$ denote the  distance traveled between locations $r$ and
$k$ at time-step $t$, as reported by the OD data.
For the same admissible paths used above, we keep track of mean path lengths
to each node $k$.
Define
\[
\mathbf{y}^{i,j,t}=[y^{i,j,t}_k],
\]
where $y^{i,j,t}_k$ is the mean distance traveled by admissible paths from $j$
to $k$ that arrive at time-step $t$, and take $y^{i,j,0}_k = 0$.

For $t \ge 1$, distances propagate via the probability-weighted update
\begin{equation}
y^{i,j,t}_k
=
\frac{1}{p^{i,j,t}_k}
\sum_{r \,:\, r \ne i,j}
\big(d^t_{kr} + y^{i,j,t-1}_r\big)\, m^t_{kr}\, p^{i,j,t-1}_r,
\label{eq:ydist}
\end{equation}
whenever $p^{i,j,t}_k>0$.

We can also organize this computation as matrix operations.
Define a matrix
\[
\tilde D^{i,j,t} = [\tilde d^{i,j,t}_{kr}],
\]
where
\[
\tilde d^{i,j,t}_{kr}
=
\frac{d^t_{kr} + y^{i,j,t-1}_r}{p^{i,j,t}_k},
\]
whenever $p^{i,j,t}_k>0$.
Then the distance update can be written as
\[
\mathbf{y}^{i,j,t}
=
\big[\,\tilde D^{i,j,t} \cdot M^t\,\big]\,
\tilde{\mathbf{p}}^{i,j,t-1},
\]
where $\cdot$ denotes element-by-element multiplication, followed by the usual
matrix product.

Again simplifying notation, let
\begin{equation*}
x^t_{ij} = y^{i,j,t}_i
\end{equation*}
denote the mean distance traveled by paths that start at $j$ and first reach $i$ at time-step $t$, without intermediate returns to $j$.

\subsection{Time-elapsed averaging}

The mean time-elapsed distance traveled from $j$ to $i$ over a window
$t_1 \le t \le t_2$ is defined by
\begin{equation}
\bar{x}^{t_1,t_2}_{ij}
=
\frac{\sum_{t=t_1}^{t_2} x^t_{ij}\,\pi^t_{ij}}
     {\sum_{t=t_1}^{t_2} \pi^t_{ij}},
\label{eq:xbar}
\end{equation}
which matches Eq.~\eqref{eq:tedist} in the main text.

Replacing step distances $d^t_{kr}$ by step travel times yields the analogous
time-elapsed travel-time quantity.

\subsection{Remarks}

\begin{itemize}
\item The exclusion of intermediate visits to $i$ and $j$ prevents trivial
self-loops from biasing the estimate.
\item The construction applies equally to travel time by replacing $d^t_{kr}$
with single-step travel times.
\end{itemize}

\section{Conceptual approaches to mitigating mobility aliasing} \label{appdx:mobaliasalg}
 
This appendix outlines a conceptual direction for mitigating mobility-aliasing
rather than a method used in the present analysis.
The time-resolution of data could be improved by changing the intervals of data aggregation to be shorter, commensurate with a statistically useful measure of speed, perhaps a standard deviation above the mean speed.  Suppose that speed is known and translates to an interval time of $T$ minutes. The aggregation would record  the trips that start in an interval $mT$ to $(m+1)T$ and end in an interval $nT$ to $(n+1)T$. For each $m$ and origin there will only be finitely many $n$ and destinations. The  record would thus be indexed by the OD pair and $m$ and $n$ (equivalently $mT$ and $nT$), a quadruple of indices, and include as data quantities of interest like trip-count, the mean (or median)  time and distance traveled. The mean/median for the  traveled time would likely be longer than the interval and those for distance could be shorter or longer  than the spatial resolution of the geohash.   This approach, while not eliminating mobility-aliasing, would help reduce it.

In the present case, the time-resolution, as noted, is coarser than needed, and potentially leads to aliasing in user and trip counts. To account for the aliasing, we could try the following idea. Continuing with the example of 3-hour, GH5, we assume that the dynamics of movements are slowly changing, at 3-hour intervals. For each coarse time-step, we need to insert $p$ transition matrices, one for each finer interval, such that their product is the original transition matrix. Denote by $H_k$ the finer-resolution transition matrices where $k=1,\ldots,p$. Omitting the superscript $t$ for the original time-step, and writing $M=M^t$,
\begin{equation*}
M=\prod^p_{k=1} H_k.
\end{equation*}
A more stringent version of this is if all the finer resolution matrices are assumed to be the same, i.e., $H_k =  H$. Then
\begin{equation*}
M=H^p.
\end{equation*}
Such a stochastic matrix $H$ is called the $p$-th root of a stochastic matrix $M$. It might not exist, however, and finding one is a challenge~\cite{hl:oprosm}.

An approximation might be attempted in finding the $p$-th root, in some distance of distance-like measure. Let the measure be $K$. The optimal approximation to the root would correspond to the minimum distance achievable over the space of stochastic matrices. Let the minimum distance be $K_{\text{min}}$:
 \begin{equation*}
K_{\text{min}} = \min_{B : B {\text{ stochastic}}}K(M,B^p).
\end{equation*}
Then the optimal approximation is $\hat H $ such that:
 \begin{equation*}
K(M,{\hat H}^p) = K_{\text{min}}.
\end{equation*}
Among such  measures could conceivably be the Relative Entropy (Kullback-Leibler divergence) or the Frobenius distance~\cite{dd:eod}. The authors have not been able to find fast converging algorithms for the approximation for the size of stochastic matrices in this paper.
 
In summary, we would need to iterate $p$ times the approximate stochastic $p$-th root of the 3-hourly transition matrix, where $p$ is the number of shorter steps in the 3-hour interval. For instance, $p= 4$ if the shorter interval is 45 minutes. The reason for choosing the approximate $p$-th root is so that the transition probabilities at the  3-hour interval are consistent with it.  We expect it to yield   different estimates for the distance and time measures that we have developed than those using the 3-hourly transition matrix, for the reason that mean trip-distance and trip-times are additive over path segments. Developing and analyzing such approximation schemes rigorously is a natural direction for future work.

\section{Supplementary diagnostics for effective distance}
\label{app:tedist-timesweep}

\subsection{Time sweep of high effective-distance OD pairs}

We examine temporal recurrence of high effective-distance events using time
sweeps of GUPs.
These visualizations are intended as diagnostics rather than inferential
results.

\noindent\emph{Synthetic data:}

\begin{figure}[H]
\centering
\includegraphics[width=\textwidth]{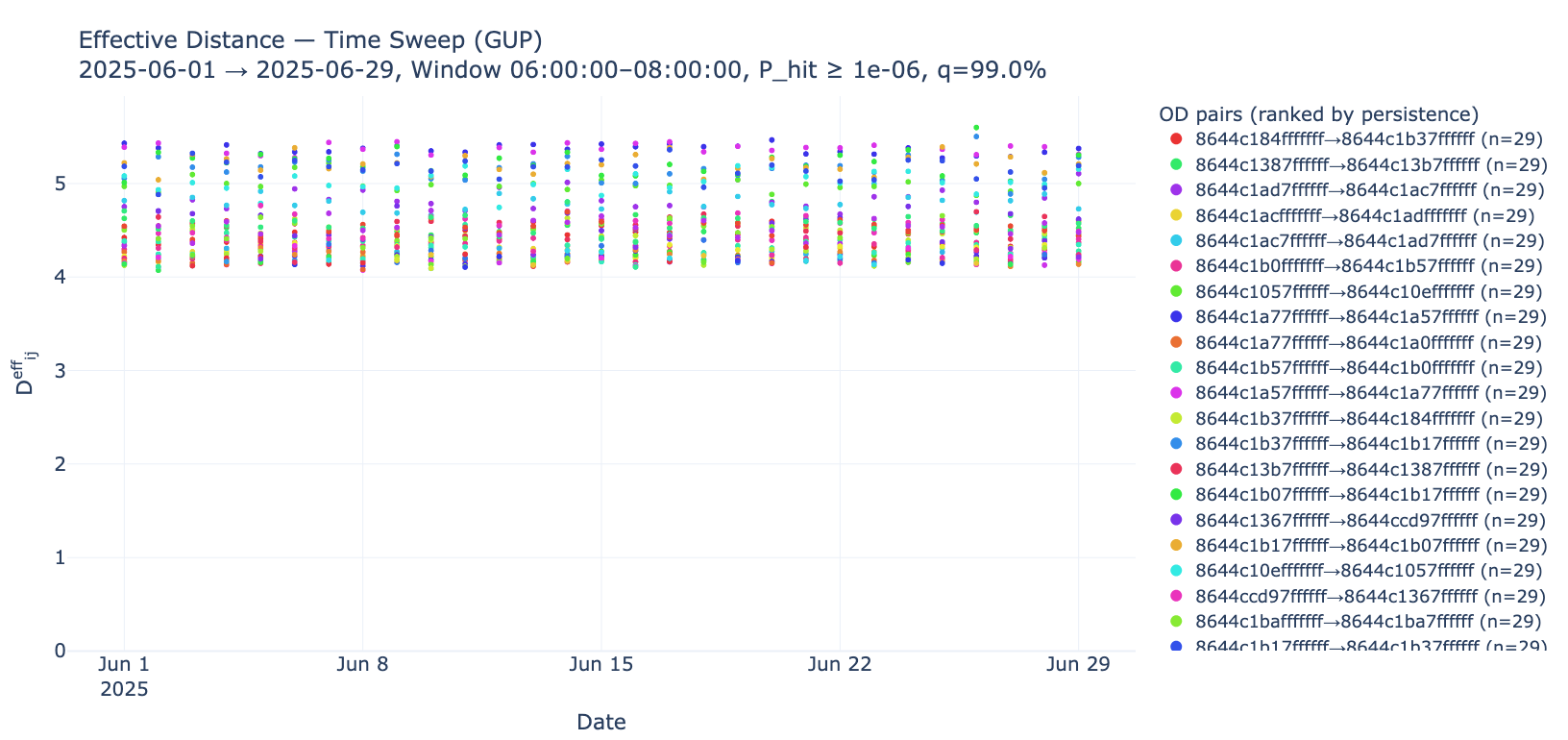}
\caption{
Synthetic time sweep of effective distances for GUPs. 
OD pairs recur with nearly uniform frequency across days, while effective
distance magnitudes exhibit bounded variability due to stochastic path
realizations within a periodic transition structure.
}
\label{fig:tedist-timesweep-synth}
\end{figure}

Figure~\ref{fig:tedist-timesweep-synth} shows a time sweep of effective
distances for synthetic GUPs over the period
2025--06--01 to 2025--06--29, for the time-elapsed window
06:00--08:00 (four 30-minute steps).
OD pairs shown fall within the top $99^{\text{th}}$ percentile of effective distance on
each day. To suppress contributions from extremely unlikely paths that arise purely from
stochastic noise, we apply a probability cutoff before computing percentile
bands.
Specifically, OD pairs are retained only if the total path-hit probability
satisfies
\[
P^{t_1,t_2}_{ij} = \sum_{t=t_1}^{t_2} \pi^t_{ij} \ge 10^{-6}.
\]
This threshold is chosen arbitrarily and is meant to remove numerically negligible connections while preserving all
structurally meaningful high effective-distance events.

The synthetic dynamics exhibit a largely regular temporal structure, but with
non-negligible variability in effective-distance magnitudes across days.
The same set of OD pairs appears persistently throughout the period, and
recurrence frequencies are highly uniform across pairs.
However, effective distances for a given OD pair are no longer perfectly
constant in time.

This variability arises from intrinsic stochastic exploration of the network
under the Markov dynamics, compounded by finite-sample effects in realized
trajectory ensembles within each time bin.
Although the underlying transition kernels are strictly periodic and admit a
periodic fixed point at the population level, effective distances are computed
from realized path ensembles.
Distance sampling within resolution-dependent envelopes introduces controlled
fluctuations in path lengths, while preserving the overall ranking and temporal
persistence of high effective-distance OD pairs.

As a result, the synthetic time sweep reflects a periodic dynamical regime with
structured stochastic variability: recurrence is highly regular, while
effective-distance magnitudes exhibit bounded fluctuations driven by
trajectory-level randomness rather than changes in the transition dynamics.

\noindent\emph{Aggregated data:}

\begin{figure}[H]
\centering
\includegraphics[width=\textwidth]{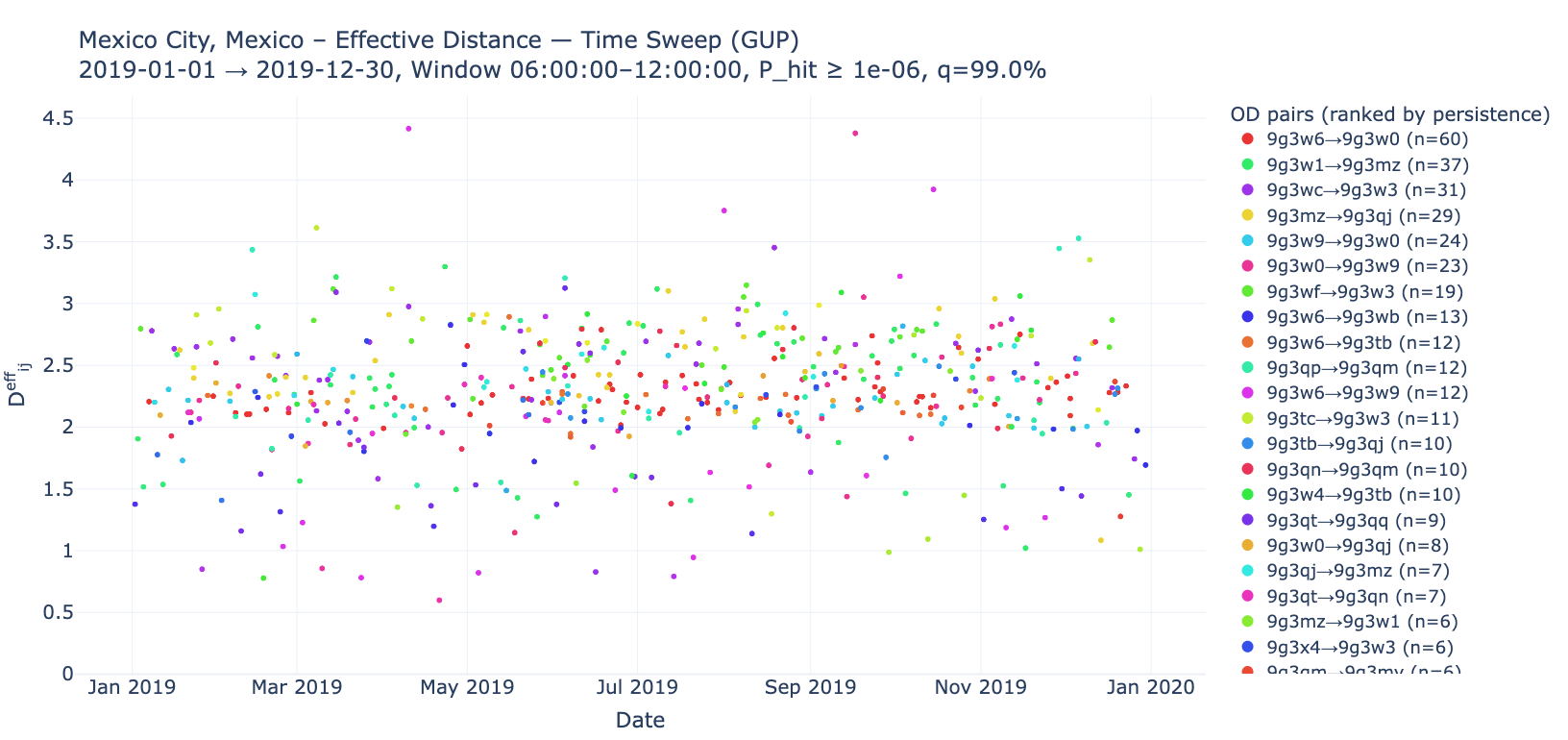}
\caption{
Mexico City time sweep of effective distances for GUPs.
 Both recurrence frequencies and effective-distance magnitudes vary
substantially across OD pairs and across days, reflecting temporal aggregation
and path ambiguity.
}
\label{fig:tedist-timesweep-agg}
\end{figure}

Figure~\ref{fig:tedist-timesweep-agg} shows the corresponding time sweep for
Mexico City aggregated OD data over 2019, for the time-elapsed interval
06:00--09:00--12:00.
As in the synthetic case, OD pairs shown fall within the top $99^{\text{th}}$ percentile
on each day. Again, we apply a probability cutoff before computing percentile
bands.
Only those OD pairs are retained whose total path-hit probability
satisfies
\[
P^{t_1,t_2}_{ij} = \sum_{t=t_1}^{t_2} \pi^t_{ij} \ge 10^{-6}.
\]
This cutoff is identical to that used in the synthetic case and serves the same diagnostic purpose.

In contrast to the synthetic setting, recurrence and effective-distance
magnitudes are both highly heterogeneous.
Some OD pairs appear frequently, while others occur sporadically, and effective
distances for a given OD pair vary substantially across days.

Several factors contribute to this irregularity.
First, the aggregated data are observed at a coarse temporal resolution
(three-hour intervals), within which multiple distinct movement patterns may be
collapsed into a single OD observation.
Second, baseline distances are estimated using median trip lengths rather than
means, reflecting reporting constraints in the data.
Third, multi-path ambiguity is unavoidable: multiple plausible intermediate
routes may contribute to a single effective-distance estimate.

\noindent\emph{Interpretation:}

The contrast between the two settings highlights the role of temporal
aggregation and stochastic structure in shaping effective-distance dynamics.
In the synthetic model, variability in effective distances arises solely from
trajectory-level sampling within a fixed, periodic transition process.
This produces bounded fluctuations around stable, recurrent OD pairs.

In the aggregated data, by contrast, variability reflects a
superposition of heterogeneous behaviors, including weekday--weekend effects,
irregular events, seasonal patterns, and context-dependent travel, all observed
at relatively coarse temporal resolution.
As a result, neither the recurrence of high effective-distance OD pairs nor
their effective-distance magnitudes remain stable over time.

While we do not perform a formal persistence or multiscale analysis, the repeated
appearance of certain OD pairs with high effective distance across days and
overlapping time windows suggests that such measures may exhibit structure
across temporal scales, a topic of interest in human mobility research~\cite{aal:shm}.
This observation motivates future work using persistence-based 
approaches to characterize such structures more systematically~\cite{ps:phsbsg,cl:dacbdccff}.

\end{document}